\documentclass[revtex4]{emulateapj}
\usepackage[]{graphicx}
\usepackage{subfigure}
\usepackage{aas_macros}
\usepackage{natbib}
\usepackage{amsmath}
\usepackage{hyperref}

\newcommand{\hst}{\emph{HST}}

\def\CIVdblt{{\rm C~}\kern 0.1em{\sc iv}~$\lambda\lambda 1548, 1550$}
\def\MgIIdblt{{\rm Mg~}\kern 0.1em{\sc ii}~$\lambda\lambda 2796, 2803$}
\def\NVdblt{{\rm N}\kern 0.1em{\sc v}~$\lambda\lambda 1238, 1242$}  
\def\OVIdblt{{\rm O}\kern 0.1em{\sc vi}~$\lambda\lambda 1031, 1037$}
\def\SiIVdblt{{\rm Si~}\kern 0.1em{\sc iv}~$\lambda\lambda1394, 1403$}
\def\AlIIIdblt{{\rm Al~}\kern 0.1em{\sc iii}~$\lambda\lambda1855,1863$}
\def\FeIIdblt{{\rm Fe~}\kern 0.1em{\sc ii}~$\lambda\lambda 2383, 2600$}

\def\AlII{\hbox{{\rm Al~}\kern 0.1em{\sc ii}}}
\def\AlIII{\hbox{{\rm Al~}\kern 0.1em{\sc iii}}}
\def\CaI{\hbox{{\rm Ca}\kern 0.1em{\sc i}}}
\def\CaII{\hbox{{\rm Ca}\kern 0.1em{\sc ii}}}

\def\CrII{\hbox{{\rm Cr~}\kern 0.1em{\sc ii}}}
\def\CII{\hbox{{\rm C~}\kern 0.1em{\sc ii}}}
\def\CIII{\hbox{{\rm C~}\kern 0.1em{\sc iii}}}
\def\CIV{\hbox{{\rm C~}\kern 0.1em{\sc iv}}}
\def\CV{\hbox{{\rm C}\kern 0.1em{\sc v}}}
\def\H{\hbox{{\rm H}}}
\def\HI{\hbox{{\rm H~}\kern 0.1em{\sc i}}}
\def\HII{\hbox{{\rm H~}\kern 0.1em{\sc ii}}}

\def\Lya{\hbox{{\rm Ly}\kern 0.1em$\alpha$}}
\def\Lyb{\hbox{{\rm Ly}\kern 0.1em$\beta$}}
\def\Lyg{\hbox{{\rm Ly}\kern 0.1em$\gamma$}}
\def\Lyfive{\hbox{{\rm Ly}\kern 0.1em$5$}}
\def\Lysix{\hbox{{\rm Ly}\kern 0.1em$6$}}
\def\Lyseven{\hbox{{\rm Ly}\kern 0.1em$7$}}
\def\Lyeight{\hbox{{\rm Ly}\kern 0.1em$8$}}
\def\Lynine{\hbox{{\rm Ly}\kern 0.1em$9$}}
\def\Lyten{\hbox{{\rm Ly}\kern 0.1em$10$}}
\def\HeI{\hbox{{\rm He}\kern 0.1em{\sc i}}}
\def\HeII{\hbox{{\rm He}\kern 0.1em{\sc ii}}}
\def\FeI{\hbox{{\rm Fe~}\kern 0.1em{\sc i}}}
\def\FeII{\hbox{{\rm Fe~}\kern 0.1em{\sc ii}}}
\def\FeIII{\hbox{{\rm Fe~}\kern 0.1em{\sc iii}}}
\def\MnII{\hbox{{\rm Mn}\kern 0.1em{\sc ii}}}
\def\MgI{\hbox{{\rm Mg~}\kern 0.1em{\sc i}}}
\def\MgII{\hbox{{\rm Mg~}\kern 0.1em{\sc ii}}}
\def\MgIII{\hbox{{\rm Mg~}\kern 0.1em{\sc iii}}}
\def\MgIV{\hbox{{\rm Mg~}\kern 0.1em{\sc iv}}}
\def\NaI{\hbox{{\rm Na}\kern 0.1em{\sc i}}}
\def\NV{\hbox{{\rm N}\kern 0.1em{\sc v}}}
\def\NII{\hbox{{\rm N}\kern 0.1em{\sc ii}}}
\def\NIII{\hbox{{\rm N}\kern 0.1em{\sc iii}}}
\def\NiI{\hbox{{\rm Ni~}\kern 0.1em{\sc i}}}
\def\NiII{\hbox{{\rm Ni~}\kern 0.1em{\sc ii}}}
\def\OVI{\hbox{{\rm O}\kern 0.1em{\sc vi}}}
\def\OI{\hbox{{\rm O}\kern 0.1em{\sc i}}}
\def\OII{\hbox{[{\rm O}\kern 0.1em{\sc ii}]}}
\def\SiII{\hbox{{\rm Si~}\kern 0.1em{\sc ii}}}
\def\SiIII{\hbox{{\rm Si~}\kern 0.1em{\sc iii}}}
\def\SiIV{\hbox{{\rm Si~}\kern 0.1em{\sc iv}}}
\def\SII{\hbox{{\rm S}\kern 0.1em{\sc ii}}}
\def\SIII{\hbox{{\rm S}\kern 0.1em{\sc iii}}}
\def\SIV{\hbox{{\rm S}\kern 0.1em{\sc iv}}}
\def\TiII{\hbox{{\rm Ti}\kern 0.1em{\sc ii}}}
\def\ZnII{\hbox{{\rm Zn~}\kern 0.1em{\sc ii}}}

\def\simlt{\mathrel{\hbox{\rlap{\hbox{\lower4pt\hbox{$\sim$}}}\hbox{$<$}}}}
\def\simgt{\mathrel{\hbox{\rlap{\hbox{\lower4pt\hbox{$\sim$}}}\hbox{$>$}}}}

\newcommand{\nh}{\mbox{$N_{\rm HI}$}} 

\newcommand{\nhi}{$N_{\rm HI}$}

\newcommand{\msun}{M_{\odot}}

\begin{document}

\title{Exploring Damped Lyman-$\alpha$  System Host Galaxies using Gamma-ray Bursts}
\author{
Vicki~L.~Toy\altaffilmark{1}, 
Antonino~Cucchiara\altaffilmark{2}$^{,}$\altaffilmark{3}$^{,}$\altaffilmark{4}, 
Sylvain~Veilleux\altaffilmark{1}$^{,}$\altaffilmark{5}, 
Michele~Fumagalli\altaffilmark{6}, 
Marc~Rafelski\altaffilmark{2}$^{,}$\altaffilmark{3}$^{,}$\altaffilmark{7}, 
Alireza~Rahmati\altaffilmark{8}, 
S.~Bradley~Cenko\altaffilmark{2}$^{,}$\altaffilmark{5}, 
John~I.~Capone\altaffilmark{1}, 
Dheeraj~R.~Pasham\altaffilmark{2}$^{,}$\altaffilmark{5}}

\altaffiltext{1} {Department of Astronomy, University of Maryland, College Park, MD 20742, USA}
\altaffiltext{2} {NASA, Goddard Space Flight Center, Greenbelt, MD 20771, USA}
\altaffiltext{3} {Space Telescope Science Institute, Baltimore, MD 21218, USA}
\altaffiltext{4} {University of the Virgin Islands, College of Science and Mathematics, 
\#2 John Brewer’s Bay, 00802 St Thomas, VI, USA }
\altaffiltext{5} {Joint Space-Science Institute, University of Maryland, College Park, MD 20742, USA}
\altaffiltext{6} {Institute for Computational Cosmology and Centre for Extragalactic Astronomy, Department of Physics, Durham University, South Road, Durham, DH1 3LE, UK}
\altaffiltext{7} {NASA Postdoctoral Program Fellow}
\altaffiltext{8} {Institute for Computational Science, University of Z\"{u}rich, Winterthurerstrasse 190, CH-8057 Z\"{u}rich, Switzerland}

\begin{abstract}
We present a sample of 45 Damped Lyman-$\alpha$ system (DLA; \nhi\ $ \geq 2 \times 10^{20} 
{\rm cm}^{-2}$) counterparts (33 detections, 12 upper limits) which host gamma-ray bursts (GRB-DLAs)
in order to investigate star-formation and metallicity within galaxies hosting DLAs. 
Our sample spans $z \sim 2-6$ and is nearly three times larger than any previously detected DLA 
counterparts survey based on quasar line-of-sight searches (QSO-DLAs). We report star formation rates (SFRs) from rest-frame UV photometry and 
SED modeling.  We find that DLA counterpart SFRs are not correlated with either redshift or HI column density. Thanks to the 
combination of \hst\ and ground-based observations, we also investigate DLA host star-formation efficiency.  
Our GRB-DLA counterpart sample spans both higher efficiency and low efficiency star formation regions compared to the local Kennicutt-Schmidt relation, 
local star formation laws, 
and $z \sim 3$ cosmological simulations.  We also compare the depletion times of our DLA hosts sample to 
other objects in the local Universe; our sample appears to deviate from the star formation efficiencies measured in 
local spiral and dwarf galaxies. Furthermore, we find similar efficiencies as local inner disks, SMC, and LBG outskirts.  
Finally, our enrichment time measurements show a 
spread of systems with under- and over-abundance of metals which may suggest that these systems 
had episodic star formation and a metal enrichment/depletion as a result of strong stellar feedback and/or 
metal inflow/outflow. 

\end{abstract}



\section{Introduction}
\label{sec:intro}

There are several successful methods to identify galaxies in the early Universe. For example, 
Lyman-break galaxies (LBGs; \citealt{Steidel:1996}) are found using the photometric drop-out 
technique around the Lyman-limit and have provided the first sample of $z\gtrsim8$ galaxies 
(e.g. \citealt{Bouwens:2010, Oesch:2012}). 
Lyman-$\alpha$ emitters (LAE), in which hydrogen recombines after ionization by young stars, 
are identified at the highest redshifts with deep near-infrared observing campaigns ($z\sim7.7$; 
\citealt{Hibon:2010, Tilvi:2010, Krug:2012}).  Because the Lyman-$\alpha$ (\Lya) line is less sensitive 
to the overall stellar continuum, LAEs are generally lower mass systems with negligible dust 
\citep{Gawiser:2007, Guaita:2011}. Additionally, 
mm/sub-mm observations have opened a promising way to study galaxies at $z\gtrsim1$ through CO molecular emission at high redshift \citep[e.g.][]{Daddi:2009}. These methods mainly probe the bright end of the luminosity 
function, at least at the highest redshifts, due to their strong stellar UV continuum. 

Another method to identify high-redshift galaxies, while also characterizing their chemical enrichment, 
utilizes bright background objects like high-redshift quasars (QSO), gamma-ray burst (GRB) afterglows, 
or, even more recently, extended background galaxies \citep{Cooke:2015, Mawatari:2016} 
to identify absorption-line systems. These detections depend only on the gas cross-section and therefore 
are less sensitive to the luminosity of the associated object (an observing bias that affects every 
high-redshift galaxy survey). Specifically, diffuse gaseous clouds in the Universe are primarily described 
by their neutral hydrogen column density (\nhi). Recent surveys have demonstrated that Damped 
Lyman-$\alpha$ systems (DLAs, see \citealt{Wolfe:2005}), characterized by \nhi\ $ \geq 2 \times 10^{20} 
{\rm cm}^{-2}$, contain $\geq80\%$ of the neutral gas available for star formation 
\citep{Peroux:2003, Prochaska:2005, Prochaska:2009, Noterdaeme:2009, Noterdaeme:2012b, Zafar:2013}. 
At $z=2-3$, they contain enough gas to account for a significant fraction (20-50\%) of stellar mass in 
all galaxies \citep{Storrie-Lombardi:2000,Wolfire:2003, OMeara:2007}.  Most importantly, they provide 
a powerful independent check on sophisticated models of galaxy formation which also include the effects 
of stellar and supernovae feedback \citep[e.g.][]{Bird:2014,Rahmati:2015}. 

Some suggested scenarios to explain the nature of high-redshift DLA galaxies include rapidly-rotating proto-galactic disks 
\citep{Prochaska:1997,Wolfe:1998,Genzel:2006, Forster-Schreiber:2009}, low surface brightness galaxies 
\citep{Jimenez:1999}, faint and small gas-rich dwarf galaxies \citep{Tyson:1988}, compact galaxies 
\citep{Nagamine:2007}, dwarf irregulars \citep{Dessauges-Zavadsky:2007}, or gaseous haloes of Lyman break 
galaxies \citep{Fynbo:1999, Moller:2002}.  There is a general consensus that the major contribution to the 
DLA population at $z\sim3$ comes from haloes with virial masses of $10^{10-12} 
\msun$ \citep{Cooke:2006aa,Barnes:2009,Font-Ribera:2012aa}. Also, \citet{Rahmati:2014} found that most 
DLAs at those redshifts are hosted by haloes with masses around or less than $10^{10} \msun$ 
(see top-right panel of Figure 6 in that paper) and, more recently, \citet{Srianand:2016}
suggested a predominant contribution, at high-redshift, of DLAs that are more compact than 
modern disk galaxies. 

To understand both the nature and evolution of the DLA population it becomes critical to identify and characterize the galaxies associated with DLAs, e.g. measuring their stellar mass, metallicity, size, and star-formation.  Understanding the types of galaxies DLAs represent will allow us to constrain which models better describe the DLA population.  There are thousands of DLAs identified from absorption-line studies, 
thanks to the Sloan Digital Sky Survey \citep{Eisenstein:2011} and the BOSS surveys \citep{Dawson:2013}.  
We can measure the neutral gas and metal content from absorption-lines, however, finding the DLA host 
galaxies that actually produced the identified features has been difficult, particularly at high redshift and/or at 
small impact parameters.  

Thus far there have only been 13 QSO-DLA confirmed galaxy counterparts. This small sample spans redshifts of $z\sim0.9-3.4$ and impact parameters of $\sim$$1-25$ kpc  \citep{Moller:1993, Moller:2002, 
Weatherley:2005, Fynbo:2011, Noterdaeme:2012a, Peroux:2012, Krogager:2012, Bouche:2013, 
Jorgenson:2014, Peroux:2016}. The majority of these DLA galaxies were found by taking spectra with multiple 
slit overlays.  This method has been successful but suffers from a strong bias towards small impact parameters as 
this is where most of the slits overlap. Moreover, the bright QSO precludes exploration at very small impact 
parameters.  It is difficult to quantify selection biases with this method as non-detection statistics are not 
reported.  Another interesting possibility is to use the Atacama Large Millimeter/submillimeter Array (ALMA) to map out CO in QSO-DLAs.  \citet{Neeleman:2016} successfully detected molecular emission from a galaxy along the projected background of a quasar with ALMA.

An independent method to identify host galaxies is the double-DLA method where a second DLA system along the line-of-sight of the 
QSO-DLA acts as a blue filter for the QSO \citep{OMeara:2006}.  This method has been successful in placing 
limits on star formation rates (SFRs) but has so far yielded few detections \citep{Fumagalli:2015}.

Finally, one can target DLAs that are identified {\it within GRB host galaxies} (GRB-DLAs): GRBs 
are extremely bright sources and can be seen up to $z\sim9$ \citep{Tanvir:2009, Salvaterra:2009, 
Cucchiara:2011}. Their bright afterglows enable the identification of the \Lya\ profile (which provides accurate 
\HI\ column density measurement) as well as metal lines at the same redshift of the GRB host (different with 
respect to QSOs, where the DLA is usually at lower-redshift). There are three key advantages of using GRB-
DLAs: 1) GRBs are very bright sources, providing exquisite high S/N spectra even at the highest redshifts; 2) 
the simple power-law continuum of the afterglow emission simplifies line identification and line profile fitting with 
respect to the more complex QSO underlying emission; 3) the afterglow emission fades away after a few days 
of the explosion, enabling {\it direct} imaging galaxies at small impact parameters \citep[$\lesssim 1-3$ kpc, as shown by][]{Blanchard:2016}
 which are often identified as the GRB host galaxies.  
\citet{Schulze:2012} demonstrated this method with a dedicated campaign to identify the galaxy counterparts for 
GRB-DLAs and sub-DLAs at $z = 2-3.6$.  The authors successfully detected a GRB-DLA counterpart for GRB 
070721B.

The main drawback with this method is that the transient nature of GRBs often makes it difficult to obtain spectra 
before the GRB afterglow has faded.  Consequently, it is challenging to assemble a large sample of GRB-DLAs; 
however, \citet{Cucchiara:2015} has reported a sample of 76 confirmed GRB-DLAs and GRB sub-DLAs (for 
which $\rm log\;{\nh}< 20.3$).  In the following sections we will use this sample as a starting point to identify and 
characterize the galaxy counterparts of these DLAs and sub-DLAs. Our compilation represents a factor of 
$\gtrsim3$ increase in the number of identified DLA galaxies to date. 

The paper is divided as follows: in \S2 we describe the GRB-DLA sample and how it compares to other GRB 
hosts or QSO-DLA samples, in \S3.1 we report star formation rates and stellar masses from our GRB-DLA 
counterparts and investigate if there is any correlation between SFR and either redshift or HI column density, in \S3.2 we 
examine the relationship between star formation rate surface density and HI gas surface density to try to
understand how star formation efficiency changes with redshift and metallicity and we compare our star formation 
efficiencies with galaxies in the local Universe, in \S4 we report enrichment times to understand how metals are formed in these 
counterparts, and in \S5 we summarize our results. 

Throughout this paper we assume a $\Lambda$CDM model with H$_0 = 69.6 \text{ km } \text{s}^{-1} \text{ 
Mpc}^{-1}$, $\Omega_m = 0.286$, and $\Omega_{\Lambda} = 0.714$ \citep{Bennett:2014}.  All magnitudes are 
in the AB system \citep{og83} and quoted uncertainties are 1$\sigma$ (68\%) confidence intervals unless 
otherwise noted.

\begin{figure*}[ht]
\centering
\includegraphics[width=3.5in]{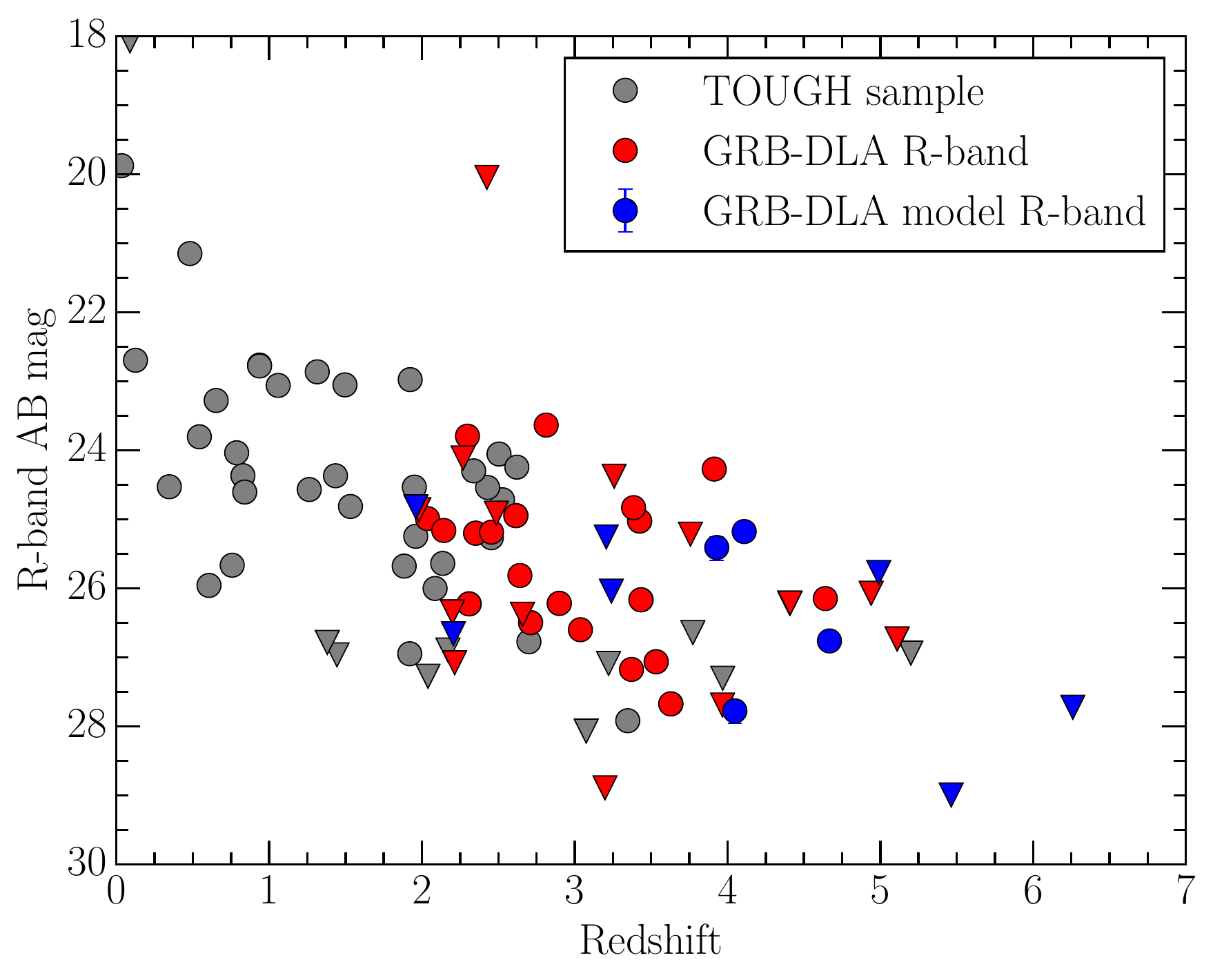}
\caption{ Distribution of R-band observations of GRB host galaxies with redshift; all data have been corrected for Galactic extinction.  Downward triangles are upper limits and circles are detections.  Red points are from R, $r^{\prime}$, F606W observations, using a flat SED to calculate R-band AB magnitude. Blue points are from using scaled SEDs from MAGPHYS (\S \ref{sec:MAGPHYS}) to determine R-band AB magnitudes (see text for details).}
\label{fig:rband}
\end{figure*}


\section{Sample and data reduction}
\label{sec:observations}

\subsection{Sample}
\label{sec:sample}

We use the GRB-DLA sample described in \citet{Cucchiara:2015} as a starting point for our search for GRB-DLA counterparts. 
This sample is comprised of 76 GRB host galaxies: 59 
confirmed GRB-DLAs and the remaining 17 objects are either GRB sub-DLAs or they only have either upper or 
lower limits on $N_{\rm HI}$ (the latter are likely sub-DLAs or Lyman limit systems).  
We conduct a literature search for photometric observations of each associated 
GRB host galaxy (see Table \ref{tab:DLAmaster} for individual observation references) and supplement these observations with data 
from the Large Monolithic Imager (LMI) on the Discovery Channel Telescope (DCT).  All of the magnitudes are 
converted to AB magnitudes using \citet{Blanton:2007} and are corrected for Galactic extinction using the dust 
map from \citet{Schlafly:2011}.  The photometry of the host galaxies is taken weeks after the GRB trigger to ensure 
that the GRB afterglow contribution is negligible. The majority of our sample is too faint to detect spectral emission lines, 
however, \cite{Blanchard:2016} performed a statistical analysis of 105 long GRBs with deep \hst\ imaging with 
1\arcsec positioning and found that 90\% of long GRBs have physical offsets of $\lesssim$5 kpc which makes 
chance associations of our sample improbable. Additionally, one expects $\lesssim$0.5 DLA \citep{Noterdaeme:2012b,Crighton:2015} 
and $\sim$1 Lyman limit system \citep{Prochaska:2010,Ribaudo:2011,OMeara:2013,Fumagalli:2013} per line-of-sight at $z=3$ which 
suggests that these are not interloping DLA or Lyman limit systems.

Out of 59 GRB-DLAs, 45 have GRB host galaxy photometric detections in at least one band or we are able to 
measure photometric limits in the rest-frame ultraviolet (UV) which directly traces star-formation. 
We do not use any photometry that is below the Lyman limit in the host 
galaxy rest-frame and our SED modeling accounts for IGM absorption (described in detail in \S \ref{sec:MAGPHYS}) 
for the three GRB-DLA and one GRB sub-DLA host galaxies that have photometric detections in the rest-frame \Lya\ forest. Throughout our 
paper we refer to these 45 GRB-DLAs as our sample (Table \ref{tab:DLAmaster}).  Our sample has a median 
$z=3.2$ and $\rm log\;{\nh} = 21.6$.  

For completeness we also include 12 sub-DLAs in Table \ref{tab:subDLAmaster}. 

\subsection{LMI data reduction}
\label{sec:lmiredux}
We use LMI to add 5 upper limits and 1 detection of DLA galaxy counterparts. 
The LMI data were detrended with a 
custom IRAF\footnote{IRAF is distributed by the National Optical Astronomy Observatories,
    which are operated by the Association of Universities for Research
    in Astronomy, Inc., under cooperative agreement with the National
    Science Foundation.} pipeline.  Individual frames were astrometrically aligned 
with {\tt Scamp} \citep{Bertin:2006} and coadded using {\tt SWarp} \citep{Bertin:2002}.  
We performed aperture photometry on the resulting coadded images using 
{\tt Sextractor} \citep{Bertin:1996} with a static 5 pixel (1.2") radius aperture, which is typical of the average seeing.  
The resulting magnitudes were 
calibrated against the Sloan Digital
Sky Survey (SDSS; \citealt{Aihara:2011}) fields.  


\subsection{Comparison to other samples}
\label{sec:compsamp}

\begin{figure*}[ht]
\centering
\includegraphics[width=3.5in]{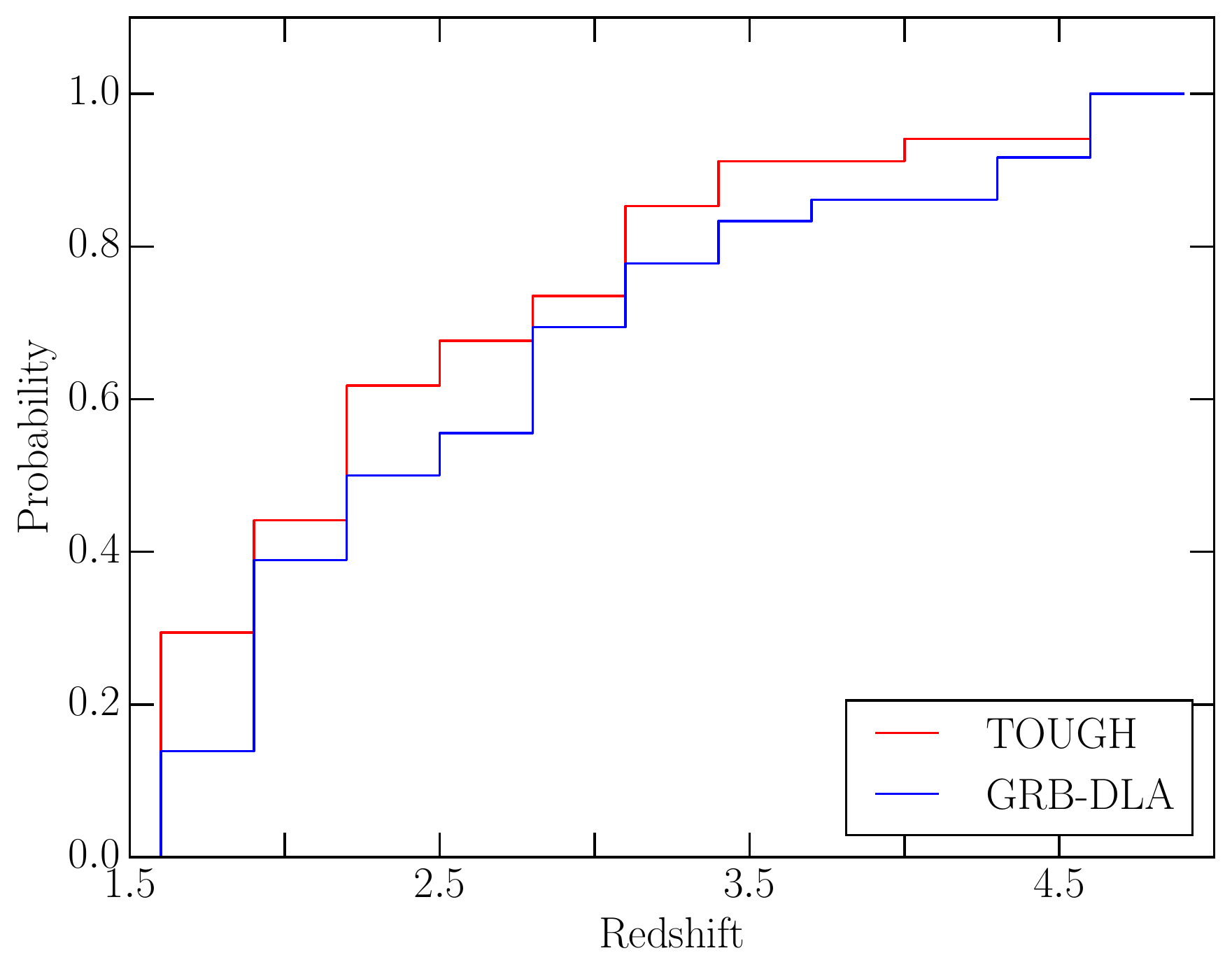}
\caption{Cumulative redshift distribution of our GRB-DLAs compared with that of the TOUGH sample. 2-sample Kolmogorov-Smirnov tests show that our sample is consistent with being drawn from the same redshift distribution as TOUGH.}
\label{fig:cumul}
\end{figure*}

We compare the observer frame R-band and redshift distribution of our sample  with The Optically 
Unbiased Gamma-ray burst Host (TOUGH) survey \citep[][see Figure \ref{fig:rband}]{Hjorth:2012}. Our DLA sample 
covers the $z \sim 2-6.3$ redshift range and a similar R-band luminosity distribution (which is usually a good proxy for 
the host rest-frame UV luminosity) as TOUGH.  In the cases where R-band is not available but we have 
$r^{\prime}$ or F606W observations, we convert to R-band assuming a flat SED between these three filters.  
Additionally, 11 GRB-DLAs do not have R-band, $r^{\prime}$-band, or F606W observations (either detections or 
limits).  For these GRB-DLAs we scale the modeled SEDs (see \S \ref{sec:MAGPHYS}) from our small sample 
of eight GRB-DLA counterparts with extensive photometric coverage to the observed magnitude and present the 
median scaled R-band value of those eight SEDs in Figure\,\ref{fig:rband}.  Note that if the 
standard deviation of the R-band value from those eight SEDs was larger than the median we report it as an 
upper limit. Also, at $z\gtrsim4$ the R-band traces flux emerging at or below the \Lya\ line ($1216$\,\AA\ rest-frame), 
therefore these values are more uncertain since they are subject to additional absorption. 

\begin{figure*}[ht]
\centering
\includegraphics[width=3.5in]{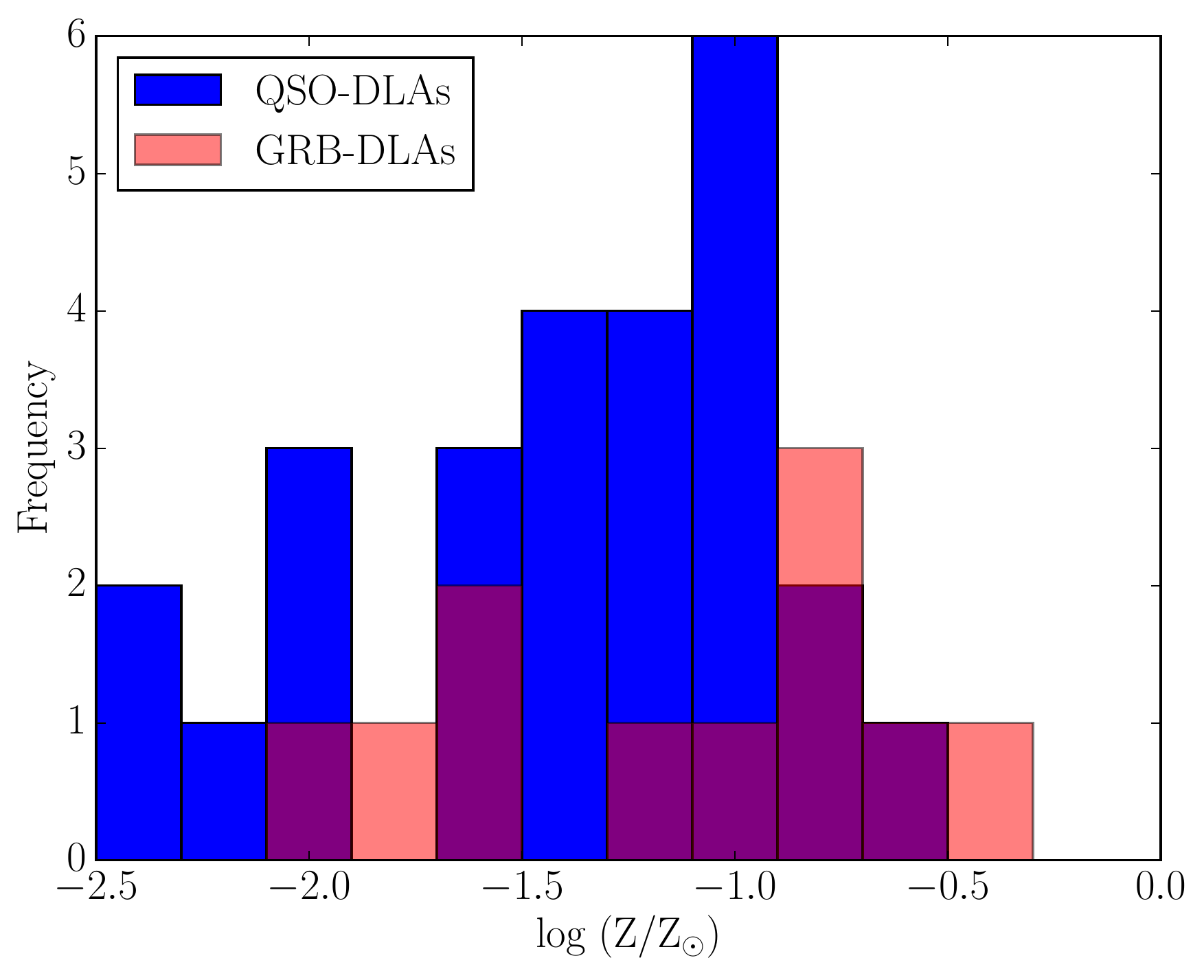}
\caption{The metallicity distribution of our sample compared with the \citet{Fumagalli:2014} double-DLA sample.  The majority of our absorption-line metallicity measurements are lower limits which we do not include in this histogram.}
\label{fig:histmet}
\end{figure*}

After we remove objects from our sample that are in the 
TOUGH survey, we run a 2-sample Kolmogorov-Smirnov test on the redshift distribution (see Figure 
\ref{fig:cumul}) over the overlapping redshift range of $z \sim 2-5$.  The $p$-value of 0.78 is consistent with 
our GRB-DLA counterpart sample and the TOUGH survey being drawn from the same GRB host population. To 
the extent that TOUGH is a representative sample of the overall GRB host population, this means that the GRB-DLAs 
hosts are also representative of the overall GRB host population. 

We also compare our sample throughout this paper to the \citet{Fumagalli:2010} sample of QSO-DLAs 
studied with the double-DLA technique which has no selection bias towards large impact parameters.  
Our sample (which covers the $\rm{N}_{HI} = 10^{20.4-22.7} \rm{cm}^{-2}$ range) 
represents an extension of the work by \citet{Fumagalli:2015}, which probes mainly lower column 
densities ($\rm{N}_{HI} = 10^{20.2-21.2} \rm{cm}^{-2}$),
providing further insights on the nature of the overall DLA counterpart population \citep[see][]{Prochaska:2007}. 
We perform a Kolmogorov-Smirnov test on the column density distribution over the overlapping column density 
range of $\rm{N}_{HI} = 10^{20.2-21.2} \rm{cm}^{-2}$ and the  $p$-value of 0.74 is consistent with our GRB-DLA 
counterpart sample and the QSO-DLA sample being drawn from the same DLA population for that range of column densities. 
However, we caution that these samples may not be from the same population for reasons discussed throughout the paper and because this $p$-value suffers from problems associated with small number statistics.

Unfortunately, it is difficult to compare our GRB-DLA metallicities with other samples because the majority of our metallicities are lower limits. Instead we only plot a histogram of our 11 GRB-DLA metallicity detections 
compared to the double-DLA sample (Figure \ref{fig:histmet}); our sample covers a similar spread in metallicity 
as the double-DLA sample with the exception of a handful of metal rich systems above ${\rm log}(Z/Z_{\odot}) > -1$. 
For more detailed analysis of our sample's metallicity distribution and a direct comparison with 
the largest compilation of QSO-DLAs to date we direct the reader to the extensive published work by
 \citet{Cucchiara:2015}, \citet{Rafelski:2012}, and \citet{Rafelski:2014}.


\section{Star Formation}
\label{sec:SF}

Star formation is correlated with the neutral gas content in a galaxy, but it is not completely clear which phase 
has a stronger causal connection with star formation: atomic, molecular, or total hydrogen \citep{Schmidt:1959, Kennicutt:1998, 
Krumholz:2009, Rafelski:2011, Elmegreen:2015, Rafelski:2016}.  Here we use atomic neutral hydrogen column 
densities measured from the damped Lyman-$\alpha$ absorption feature and assume that the molecular 
hydrogen has a negligible contribution.  This is supported by the small ($\sim1\%$) molecular hydrogen 
detection rate in a blind and uniformly selected DLA survey \citep[][]{Jorgenson:2013, Jorgenson:2014a}
and by targeted surveys \citep{Noterdaeme:2008}.

Additionally, it is rare to detect molecular absorption features in GRB afterglow spectra (supported by the few $H_2$ measurement along few GRB lines of sights, e.g., \citealt{Prochaska:2009a,Kruhler:2013,DElia:2014,Stanway:2015}) due to the unavailability of the required high-resolution instruments and blue spectral coverage.

We caution that the GRB afterglow line-of-sight is 
probing a much smaller area ($\sim$parsec scale) of the much larger galaxy ($\sim$kiloparsec scale); however, if 
GRBs occur in star-forming regions we expect them to encounter molecular hydrogen whereas the QSO may be ouside of the star-forming region.

We calculate star formation rates (SFRs) from rest-frame UV luminosities (see Section \S \ref{sec:SFR}) and 
investigate if there is any correlation with redshift or the ISM metallicities (as determined by the absorption 
features).  We then calculate star formation rate surface densities and HI surface densities to 
explore star formation efficiencies (Section \S \ref{sec:KS}), and finally we examine possible redshift and absorption 
metallicity trends in comparison with the Kennicutt-Schmidt relation at both local, $z=0$, and at higher redshifts 
(from cosmological simulations).


\subsection{Star Formation Rates}
\label{sec:SFR}

We calculate SFRs using three methods. The first and preferred method is SED modeling 
using MAGPHYS described in \S \ref{sec:MAGPHYS}.  We limit the use of SED modeling to GRB-DLA 
counterparts that have photometric detections in at least three separate bands which is the minimum for MAGPHYS to converge to a reasonable SED fit (although with large parameter errorbars in cases with few photometric points).  The second method is using single band 
detections corresponding to rest-frame UV bandpass to calculate rest-frame UV SFR (see \S 
\ref{sec:singleSFR}).  For consistency we compare SFRs based on the first two methods: SFR values from these 
two methods reasonable agree with each other usually within a factor of two, but in rare cases may vary by a factor of five most likely due to 
different accounting of dust extinction.  Generally the single band SFRs are in agreement or are slightly lower than 
those derived from SED modeling with MAGPHYS. 

If we are unable to use either of the first two methods and we have at least one detection 
in another filter, we scale the SEDs from the DLA counterparts that were fit with the first method to match the 
detected host galaxy flux.  We then use the scaled SEDs to estimate the rest-frame UV flux and use the median 
and standard deviation of the scaled SEDs to calculate the rest-frame UV SFR.  Finally, if there are no 
detections in any band but there are upper limits in the rest-frame UV band, we calculate SFR upper limits using 
the second method.

Photometric measurement were made using aperture photometry technique, using the Hubble Space Telescope 
(HST) point spread function (PSF) for GRB-DLA counterparts with HST data and the DCT 1.2" PSF for the 
ground-based data (corresponding to $\sim$2 and $\sim$17 kpc diameter apertures respectively).  The large difference in 
apertures comes from the fact that HST is able to resolve the host galaxy.  We assume that the light from 
unresolved sources is solely from the host galaxy and background sky.

All SFRs are calculated from dust-corrected observations unless otherwise stated. The host extinction, $A_V$, 
is taken either from SED models or from GRB afterglow measurements using a Small Magellanic Cloud (SMC)-
like extinction law which has been shown to best depict the GRB explosion environment \citep[e.g.][]{Schady:2012}. 

We assume the host extinction is the same as the GRB line-of-sight extinction which  
\citet{Perley:2013} has shown is fairly consistent within a factor of 2-3. If the host extinction is an upper limit, we 
use that value in all dust-corrected calculations and report SFR upper limits.  In Table \ref{tab:DLAmaster} we 
report these GRB-DLAs with SFR error estimates but treat these as SFRs upper limits in all plots using dust-corrected SFRs.
Our host extinction is in general higher than the $A_V \lesssim 0.1$ reported for DLAs in the SDSS survey for 
our sample's column densities \citep{Murphy:2016}. This may likely be because GRB-DLAs are found at smaller 
impact parameter of $\lesssim$5 kpc \citep{Blanchard:2016} than the general DLA population of $1-25$ kpc \citep{Fumagalli:2015} or more simply because our DLA sample
traces in general metal rich, and likely dust rich, systems \citep[for example, see correlation between $E(B-V)$ and metal lines equivalent widths in][]{Murphy:2016}.

\subsubsection{SED fitting Star Formation Rate}
\label{sec:MAGPHYS}

We use MAGPHYS with the {\tt HIGHZ} extension \citep{da-Cunha:2008,da-Cunha:2015}, to model the host 
galaxy SEDs from photometry.   MAGPHYS models templates to the data and returns a SED with fitted 
parameters which include SFR, stellar mass (M$_*$), dust mass (M$_{\rm dust}$), and $A_V$. This particular 
package is well suited for $z>1$ galaxies and takes into account bursty star formation which is appropriate for 
GRB host galaxies as suggested by \citet{Hunt:2014}. MAGPHYS uses a continuous model of star formation 
with superimposed random bursts that happen at equal probability at all times up to the age of the galaxy.  The 
probability is set such that 50\% of the galaxies within the library have had a burst of star formation within the 
past 2 Gyr with bursts lasting $\sim$$10^{7} - 10^{8}$ years. MAGPHYS also accounts for IGM absorption 
and uses a Gaussian distribution centered around the mean IGM effective absorption from \citet{Madau:1995} 
for each model template. 

We only select objects that have at least three photometric detections in order to break some parameter degeneracy 
and then include, if available, upper limits.  We 
have nine GRB-DLAs that fit this criterion; however, GRB 080607 returns an unconstrained SFR and M$_*$.  
This particular host galaxy has an extremely high host extinction and HI column density that is atypical of the 
majority of galaxies \citep{Wang:2012, Perley:2011, Chen:2010, Prochaska:2009a}. 

\begin{figure*}[ht]
\centering
\begin{minipage}[b]{0.48\linewidth}
	\includegraphics[width = 1\linewidth]{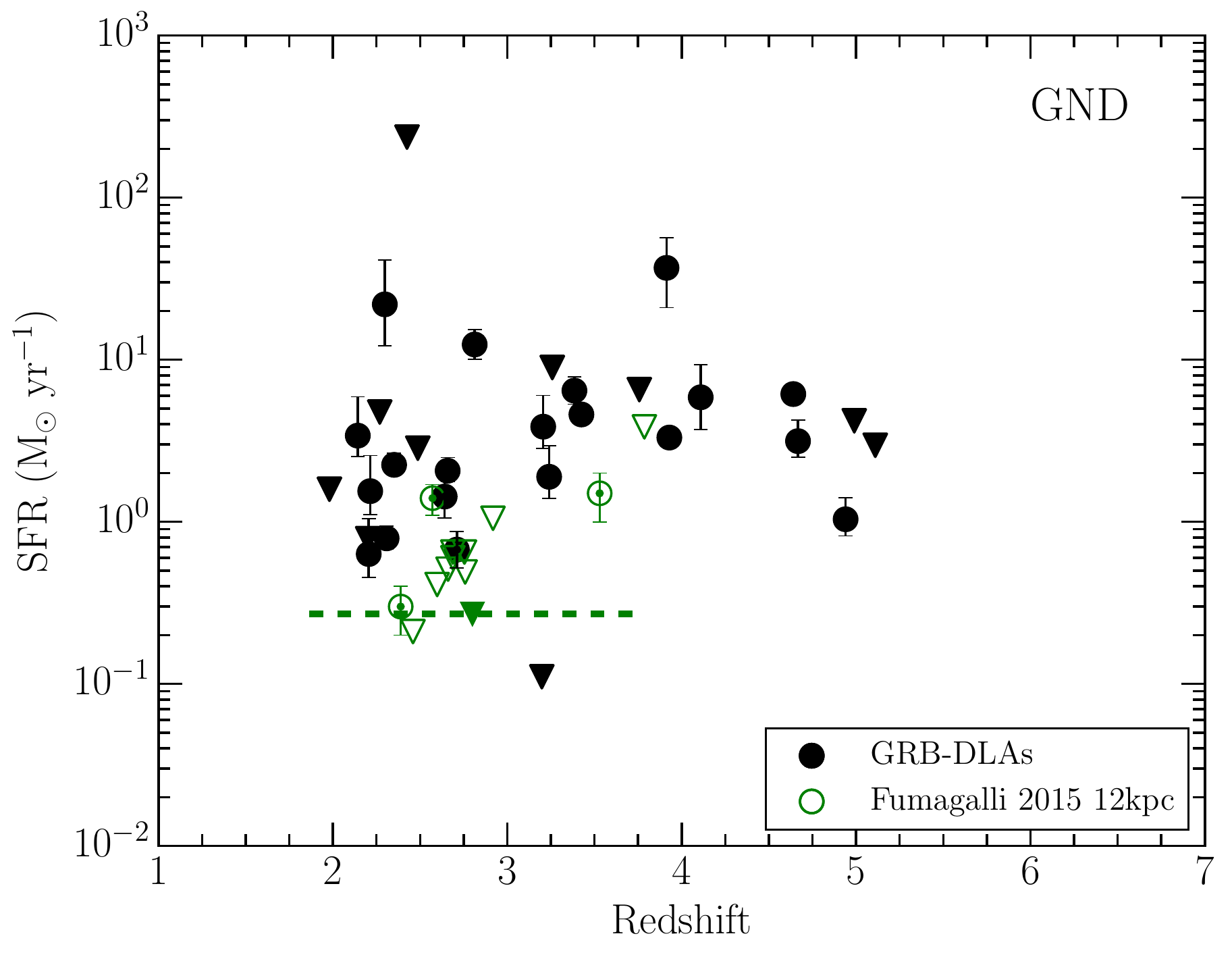}
\end{minipage}
\quad
\begin{minipage}[b]{0.48\linewidth}
	\includegraphics[width = 1\linewidth]{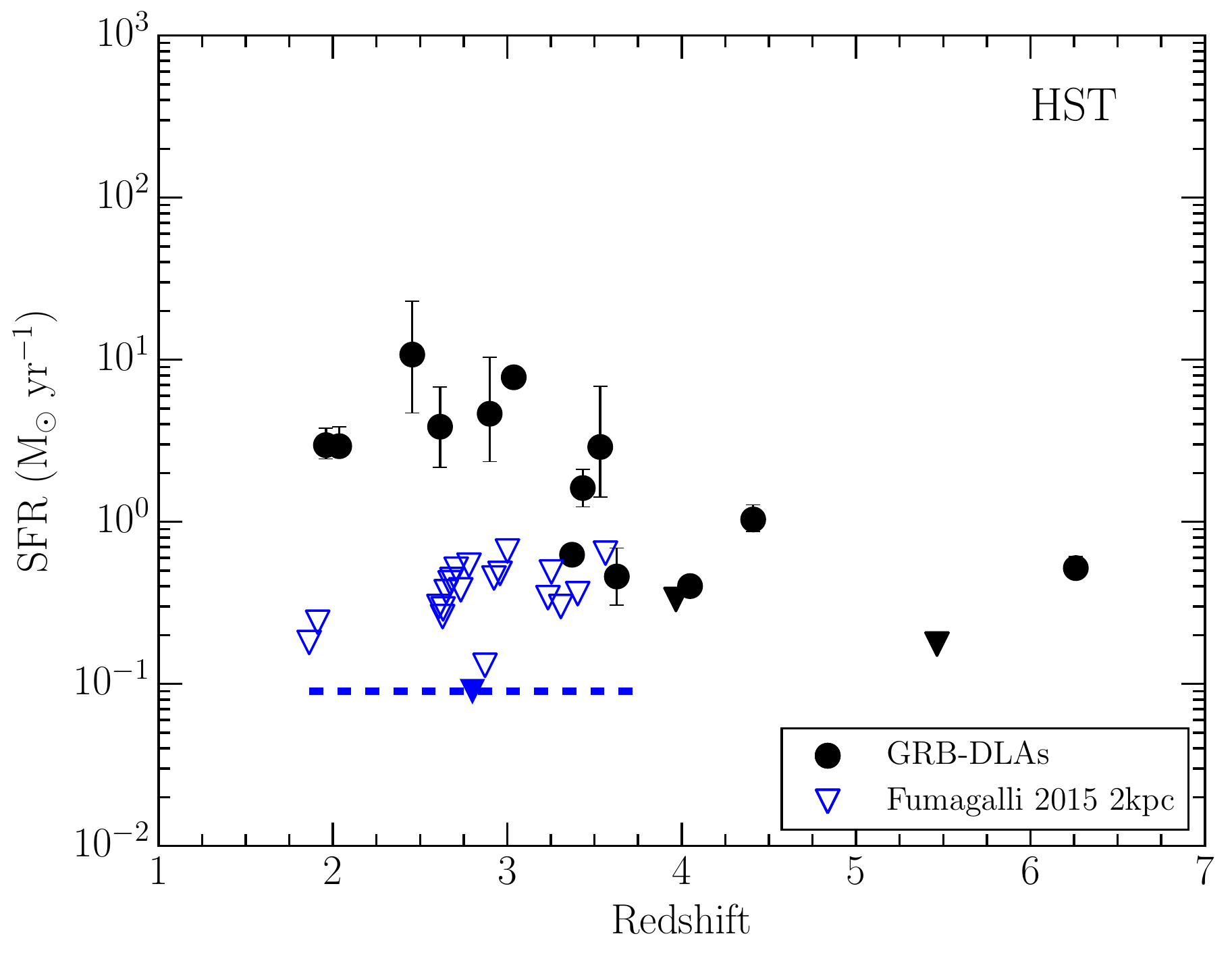}
\end{minipage}
\begin{minipage}[b]{0.48\linewidth}
	\includegraphics[width = 1\linewidth]{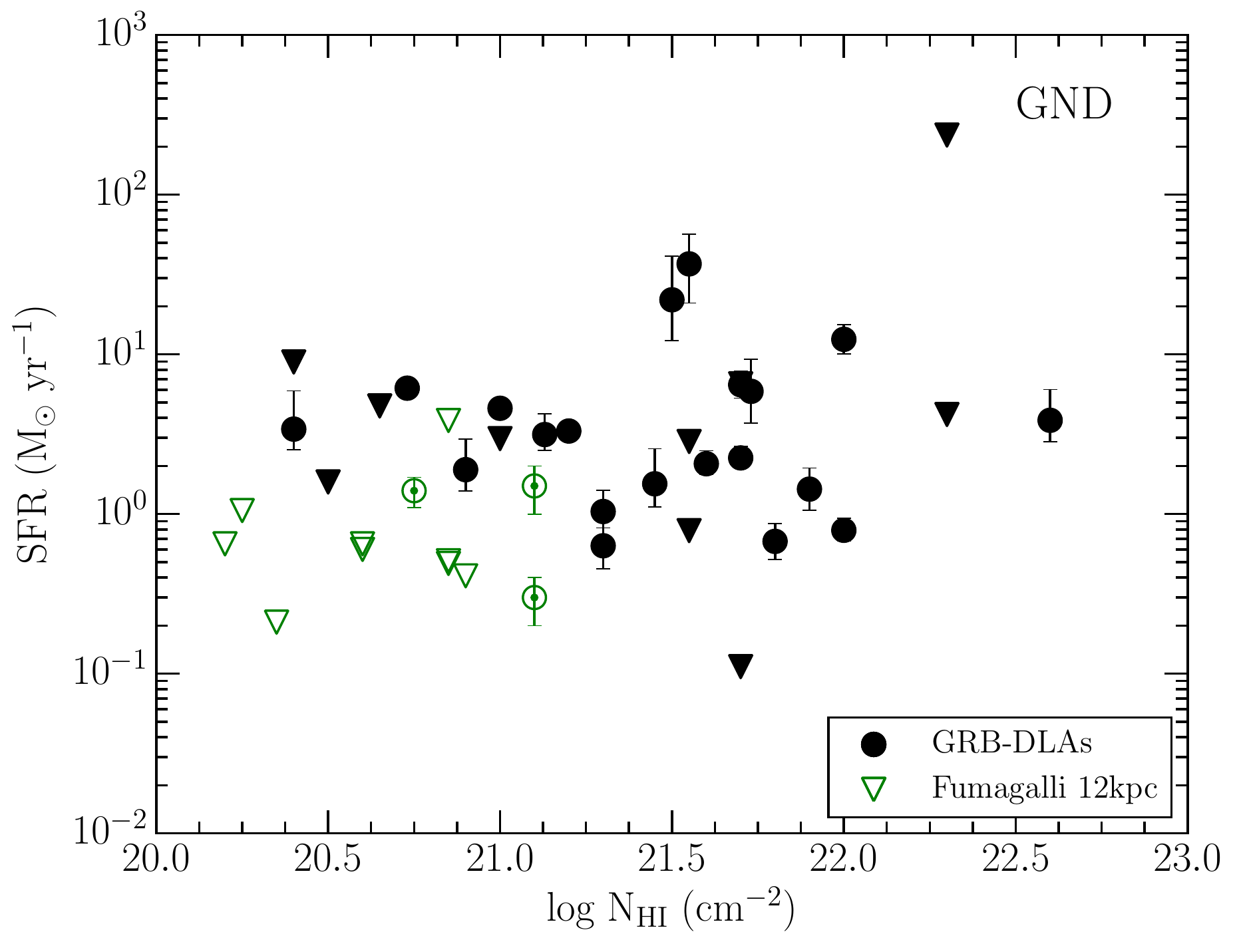}
\end{minipage}
\quad
\begin{minipage}[b]{0.48\linewidth}
	\includegraphics[width = 1\linewidth]{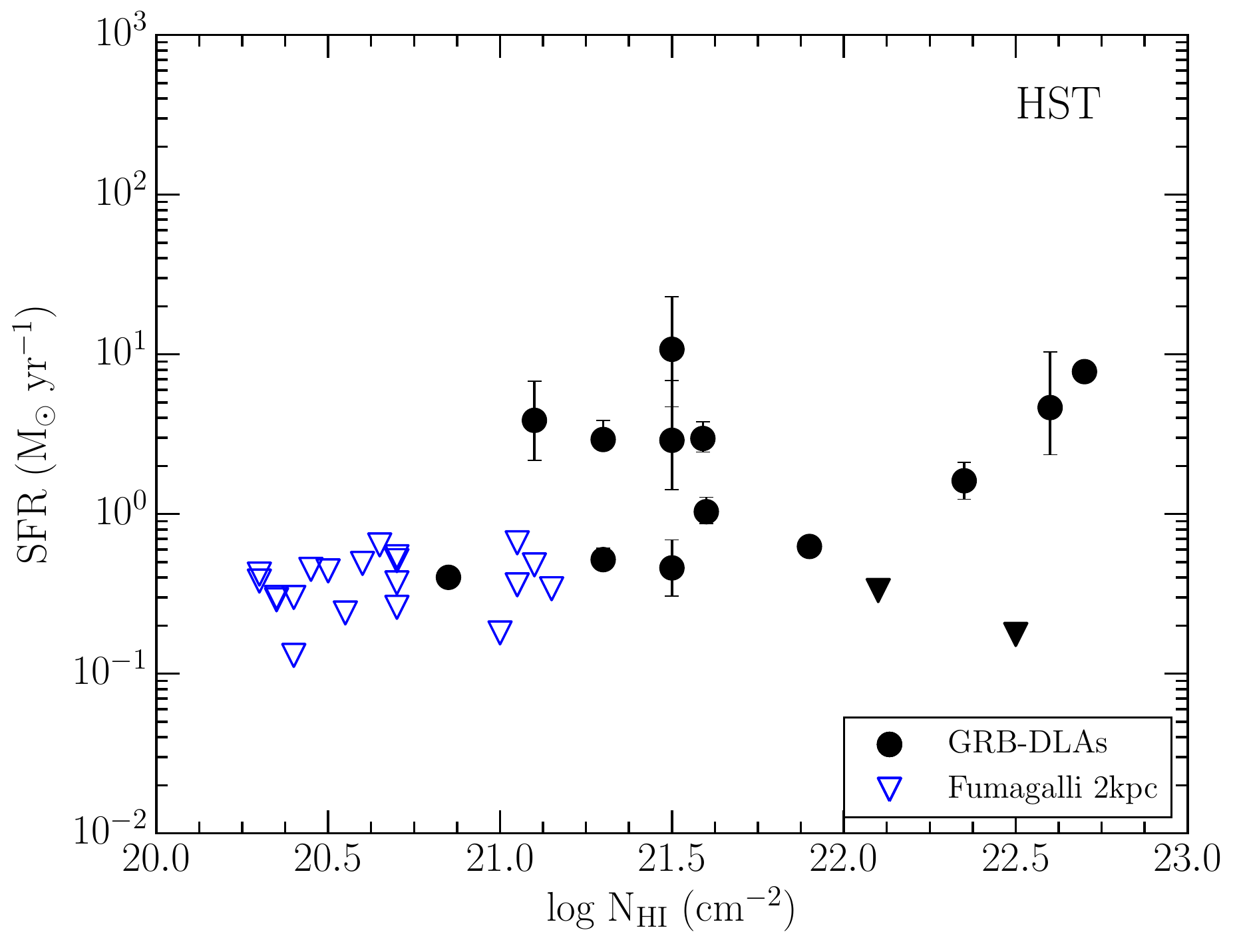}
\end{minipage}

\caption{ Comparing SFRs in our sample (black points) with \citet{Fumagalli:2015} double-DLA SFRs (green/blue unfilled points) for both ground-based and \hst\ data.  Both datasets are uncorrected for dust for direct comparison. Triangles represent upper limits. Our sample uses the DCT 1.2\arcsec PSF ($\sim$17kpc diameter) apertures for ground-based data and the \hst\ PSF ($\sim$2kpc diameter) apertures for \hst\ data. ({\it top left}) Ground-based SFRs vs. redshift.  There are three double-DLA detections, but one may be contaminated by the QSO (see \citealt{Fumagalli:2015} for details).  The dashed green line is a deep limit from a composite image. ({\it top right}) \hst\ SFRs vs. redshift.  The dashed blue line is a deep limit from a composite image. ({\it bottom left}) Ground-based SFRs vs. HI column density.  ({\it bottom right}) \hst\ SFRs vs. HI column density.}
\label{fig:sfr} 
\end{figure*}

\subsubsection{Single band UV Star Formation Rate}
\label{sec:singleSFR}

We use the relations for UV luminosities from \citet{Savaglio:2009} to determine SFR from a single photometric 
band: 

\begin{equation}
\label{eq:1500}
	\mathrm{SFR}_{1500} = 1.62 \times 10^{-40} \frac{L_{1500, \mathrm{corr}}}{\mathrm{erg s}^{-1} \mathrm{\AA}^{-1}}\, 
				\mathrm{M}_\odot \mathrm{yr}^{-1}
\end{equation}

\begin{equation}
\label{eq:2800}
	\mathrm{SFR}_{2800} = 4.33 \times 10^{-40} \frac{L_{2800, \mathrm{corr}}}{\mathrm{erg s}^{-1} \mathrm{\AA}^{-1}}\, 
				\mathrm{M}_\odot \mathrm{yr}^{-1}
\end{equation}

\begin{equation}
\label{eq:3600}
	\mathrm{SFR}_{3600} = 5.47 \times 10^{-40} \frac{L_{3600, \mathrm{corr}}}{\mathrm{erg s}^{-1} \mathrm{\AA}^{-1}}\, 
				\mathrm{M}_\odot \mathrm{yr}^{-1}
\end{equation}

\noindent Equations \ref{eq:1500}-\ref{eq:3600} were derived from samples with simultaneous H$\alpha$ and 
UV detection suitable for GRB host galaxies and are for dust-corrected rest-frame UV luminosities. In Figure 
\ref{fig:sfr} we present only dust-uncorrected rest-frame UV luminosities to directly compare with 
\citet{Fumagalli:2015}, but in all other figures and tables we present dust-corrected rest-frame UV SFRs.  
We note that other objects that we compare with in this paper use H-$\alpha$ to SFR conversions from \citet{Kennicutt:1998} (e.g. \citealt{Fumagalli:2015} and \citealt{Rafelski:2016}): direct comparison to \citet{Savaglio:2009} can result in a difference of a factor of $\lesssim$2 in SFRs (which includes factors for different initial mass functions).

To determine rest-frame UV SFR, we consider observations redward of the rest-frame \Lya\ line and from filters that have rest-frame effective 
wavelengths within 250\AA\ of 1500\AA, 2800\AA, or 3600\AA\ when we use these relations.  We have 12 
GRB-DLAs with rest-frame UV detections (four of which have $A_V$ upper limits so we list the SFRs as upper limits) and 12 GRB-DLAs with rest-frame UV limits.   

Additionally, we have another 12 GRB-DLAs that have detections redder than the rest-frame UV (one of which has an $A_V$ upper limit so we list the SFR as an upper limit).  We use the 
scaled SEDs from the eight GRB-DLAs fit with MAGPHYS (we do not include GRB 080607 in this fit for reasons 
described in \S \ref{sec:MAGPHYS}) and calculate the SFR using Eq. \ref{eq:1500}-\ref{eq:3600} for the closest 
wavelength to our rest-frame observed effective wavelength. We report the median and standard deviation SFR 
of these eight scaled SED in Table \ref{tab:DLAmaster}.  We also find that our SFR measurements are in good agreement with literature values (e.g. SHOALS sample; \citealt{Perley:2013}).

\begin{figure*}[ht]
\centering
\includegraphics[width=0.6\linewidth]{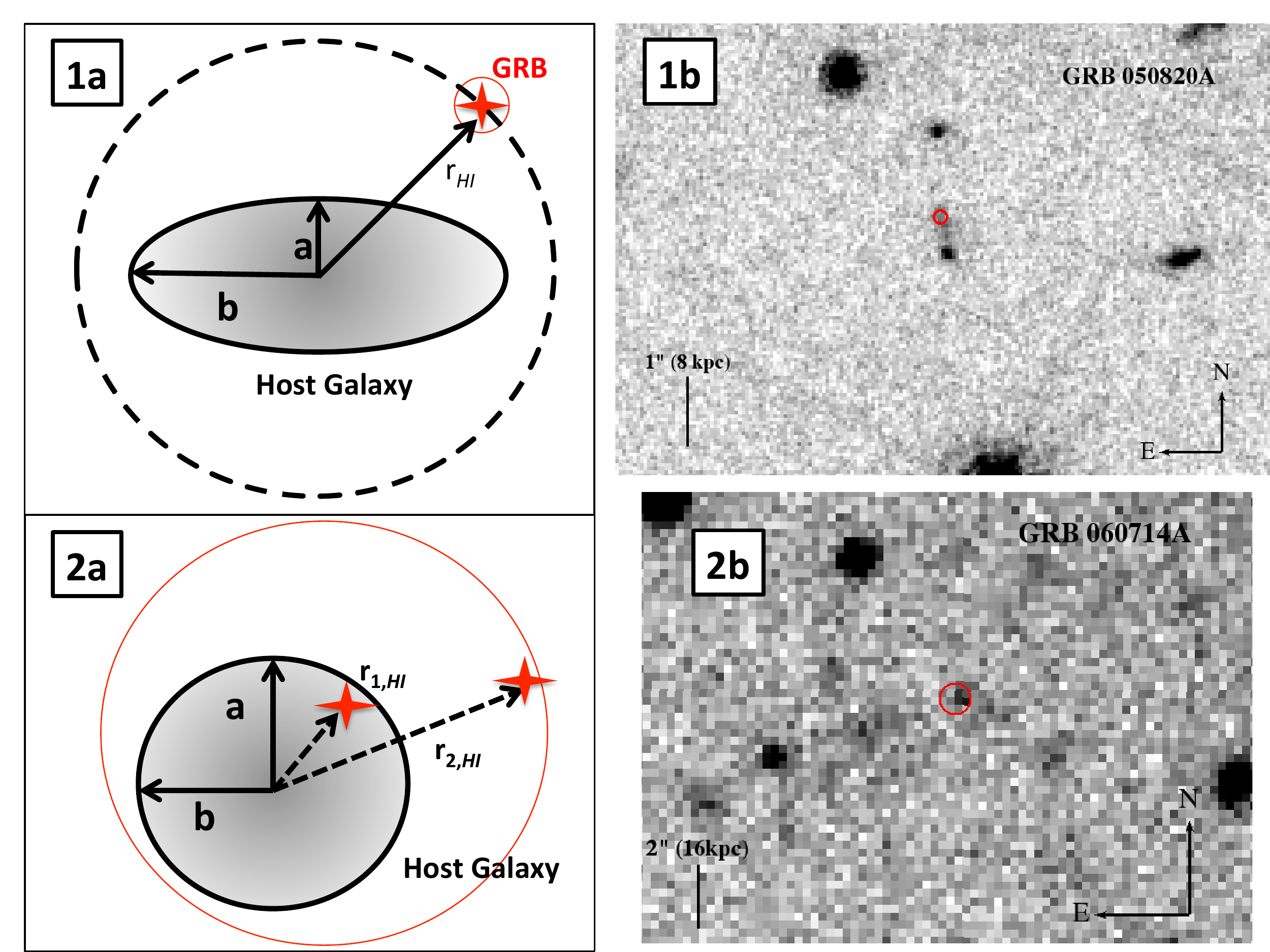}
\caption{ ({\it top left, 1a}) Ideal case where the GRB, and therefore the \H I\ gas, is extremely well localized (red circle is localization error) and can be identified relative to the host galaxy. ({\it top right, 1b}) An observed case, GRB 050820A, close to the top-left idealized configuation. The GRB-DLA is localized to sub-arcsec precision (red circle) from rapid follow-up of the afterglow with \hst and the DLA galaxy has been resolved using \hst. ({\it bottom left, 2a}) Realistic case where the GRB, and therefore the \H I\ gas, has a large error circle (red circle) that can place the GRB within the host galaxy or on the outskirts. ({\it bottom right, 2b}) An observed example, GRB 060714A, close to the bottom-left realistic configuration.  The GRB is localized to $\lesssim$1\arcsec (red circle) and, although observed with Keck, the host galaxy is unresolved.   }
\label{fig:cartoon}
\end{figure*}

\begin{figure*}[ht]
\centering
\begin{minipage}[b]{0.48\linewidth}
	\includegraphics[width = 1\linewidth]{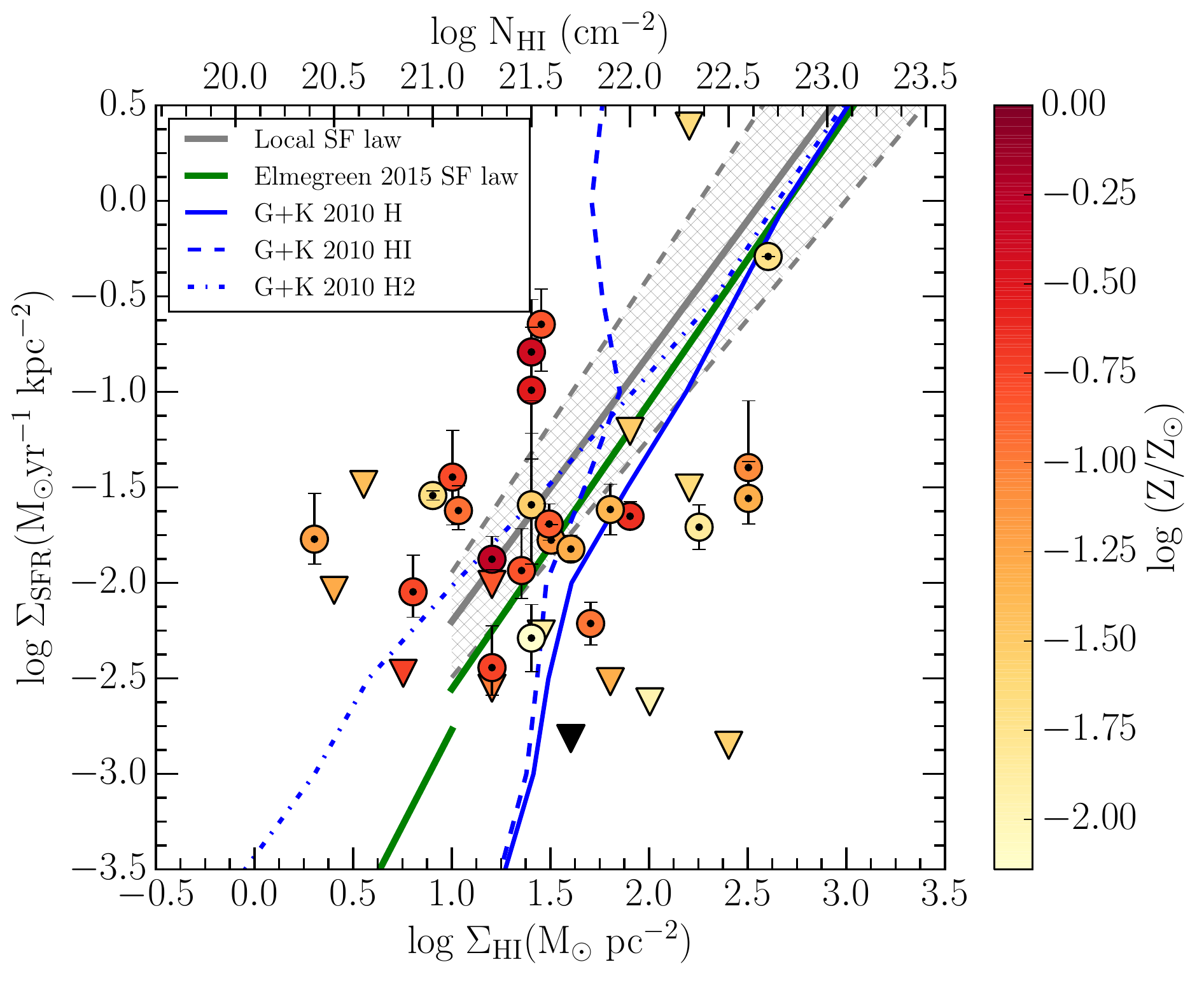}
\end{minipage}
\quad
\begin{minipage}[b]{0.48\linewidth}
	\includegraphics[width = 1\linewidth]{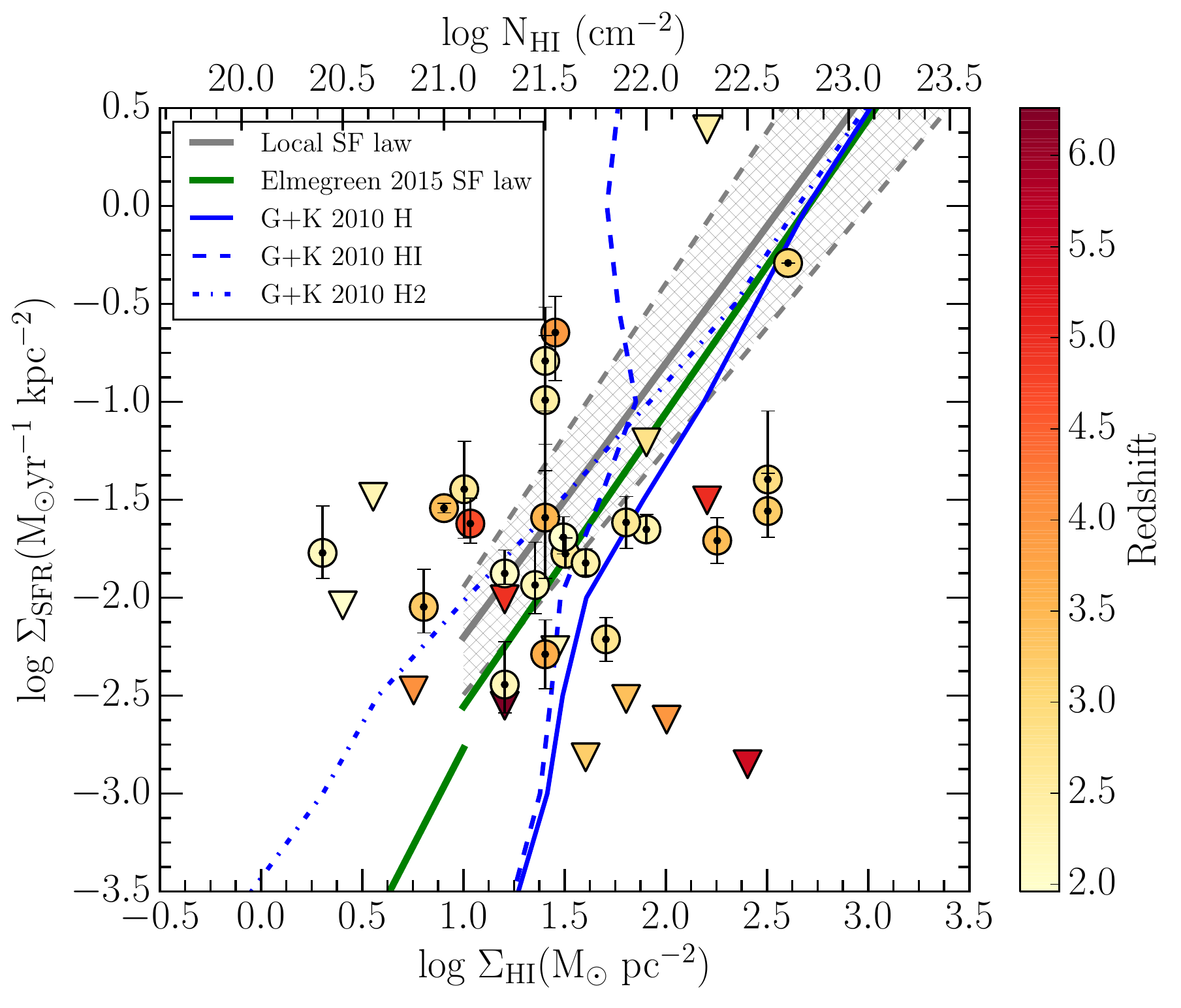}
\end{minipage} 
\caption{ Dust corrected SFR surface densities vs. HI gas surface densities of our GRB-DLA counterpart sample.  We do not include GRB-DLAs that do not have host dust extinction measurements.  Upper limits are shown as triangles as are GRB-DLAs with host extinction upper limits. We overplot the local Kennicutt-Schmidt relation with errors (gray and gray hash; \citealt{Schmidt:1959, Kennicutt:1998}), the local SF law from \citet{Elmegreen:2015} (green), and $z \sim 3$ \citet{Gnedin:2010} simulations for total hydogren, atomic, and molecular gas (blue solid, dashed, and dotted respectively).  The {\it total} neutral gas from \citet{Gnedin:2010} SF laws should be shifted to the right since we are plotting against the {\it atomic} hydrogen gas content.  Additionally the {\it molecular} hydrogen gas should be shifted to the left since we expect there to be more {\it atomic} hydrogen gas than {\it molecular} hydrogen gas.  \textit{(Left)} SFR surface densities vs. HI gas surface density color coded with metallicity; black points have no metallicity measurements from absorption lines.  \textit{(Right)} SFR surface densities vs. HI gas surface density color coded with redshift.}
\label{fig:surfacedens} 
\end{figure*}


\subsubsection{DLA host Star Formation Rates}

In Figure \ref{fig:sfr} we compare the dust-uncorrected SFRs with the dust-uncorrected SFR detections and limits derived by 
\citet{Fumagalli:2015}. Similar to this study, we also take full advantage of our large dataset and probe \emph{in 
situ} DLA counterpart SFRs within compact ($\sim$2 kpc using \hst\ data) and more extended regions 
($\sim$17 kpc using our ground-based observations). 
The majority of our sample has generally higher SFRs than the double-DLA limits, however, in some cases we obtain SFRs similar to the 
double-DLA limits both from ground-based and \hst\ observations 
(downward triangles). This result displays the effectiveness of targeting GRB-DLA counterparts: not only is our DLA 
detection rate higher than \citet{Fumagalli:2015}, but our DLAs (when we combine ground and HST data) span a 
larger range of both redshift and column densities and trace intrinsic SFR over four orders of magnitude 
($10^{-1}-10^2 $M$_{\odot}/$yr).  

Nevertheless, some DLA counterparts identified along QSOs have measured SFRs with $1-30 \rm M_{\odot} \rm yr^{-1}$ (see \citealt{Fumagalli:2015} and references within). 
It may be that it is more difficult to detect these high SFR DLAs along 
QSOs using an unbiased impact parameter survey as we mentioned in \S\ref{sec:intro} or they may be from an 
entirely different counterpart population.

We caution that SFRs of DLAs within GRB hosts may be skewed towards higher values than the general DLA 
population because our sample is taken from long-duration GRBs which are known to be associated with the evolution of 
massive stars (see \citet{Woosley:2006} for review) and are therefore associated with galaxies which have 
higher specific SFRs \citep{Japelj:2016}. 

Recent work by \citet{Perley:2016} has shown that the $z\gtrsim 2$ 
GRB host population seems to be consistent with the general cosmic star-formation rate, strengthening the idea that our 
DLA sample may be an important complement to our current understanding of the nature of DLAs. 

Also, DLA counterpart SFRs have been predicted to be higher for higher column densities and higher metallicities 
\citep{Krumholz:2009, Gnedin:2010, Rafelski:2011, Noterdaeme:2014, Rahmati:2014, Rafelski:2016}.
Our SFRs appear to be independent of 
column density in Figure \ref{fig:sfr}: the 5 detections (including both ground and HST data) with $\rm{N}_{HI} \leq 10^{21} \rm{cm}^{-2}$ have similar SFRs of those with high HI column densities and \citet{Rahmati:2014} simulations show that only 5\% of galaxies with $\rm{N}_{HI} = 10^{20-21} \rm{cm}^{-2}$ have SFRs $>10 M_{\odot} \rm yr^{-1}$. Again, we caution that SFR is a global measurement of the host counterpart 
whereas HI column density is measured along the line-of-sight of the GRB afterglow and there may be some 
scatter in the line-of-sight measurement compared to the average DLA HI column density.  Since the majority 
of our metallicity measurements are lower limits it is difficult to determine if metallicity plays an important role, if any at all,
as presented in some cosmological simulations \citep[][]{Rahmati:2016}.

We compare our distribution of SFRs within $z=2-4$ and $\rm{N}_{HI} = 10^{21.5-22} \rm{cm}^{-2}$ to 
simulation results from \citet{Rahmati:2014} at $z=3$ with the same \nhi\ range.  Our sample has a total of 15 
objects that meet these criteria and $33\%\pm8\%$ of them have SFRs $< 1 M_{\odot} \rm yr^{-1}$ where 
we assume the error is primarily poissonian. This number is slightly lower than the predicted 45\% by 
\citet{Rahmati:2014}. While the number of GRB-DLAs in this comparison is still small, future and more complete 
GRB-DLAs surveys (like the SHOALS survey) will provide more accurate tests for cosmological simulations and 
the conversion of neutral gas into stars (e.g. stellar mass).

It is also evident from our results that the DLA counterpart SFRs appear to be independent of redshift and 
our detections are all above the double-DLA upper limits for 
both the ground-based and the \hst\ observed GRB-DLAs, although the higher SFRs measured in the ground data may
be affected by unresolved part of the GRB hosts (especially at high-$z$).
In fact, as pointed out by \citet{Fumagalli:2015}, resolving the exact location of the emission of 
the DLA counterparts plays a critical role in our understanding of the DLA properties (see Figure \ref{fig:cartoon}), and 
only more \hst\ data, in combination with more accurate GRB afterglow localization will enable precise DLAs \emph{in situ} SFR measurements.
We note, for our current sample, that the probability of chance association with \hst\ are typically $\lesssim$0.05 so it is very unlikely 
that these are interloping galaxies \citep{Blanchard:2016}, but are indeed region of star-formation within the GRB host (Figure  \ref{fig:cartoon}, panels 1a and 1b).

\begin{figure*}[ht]
\centering
\includegraphics[width=4in]{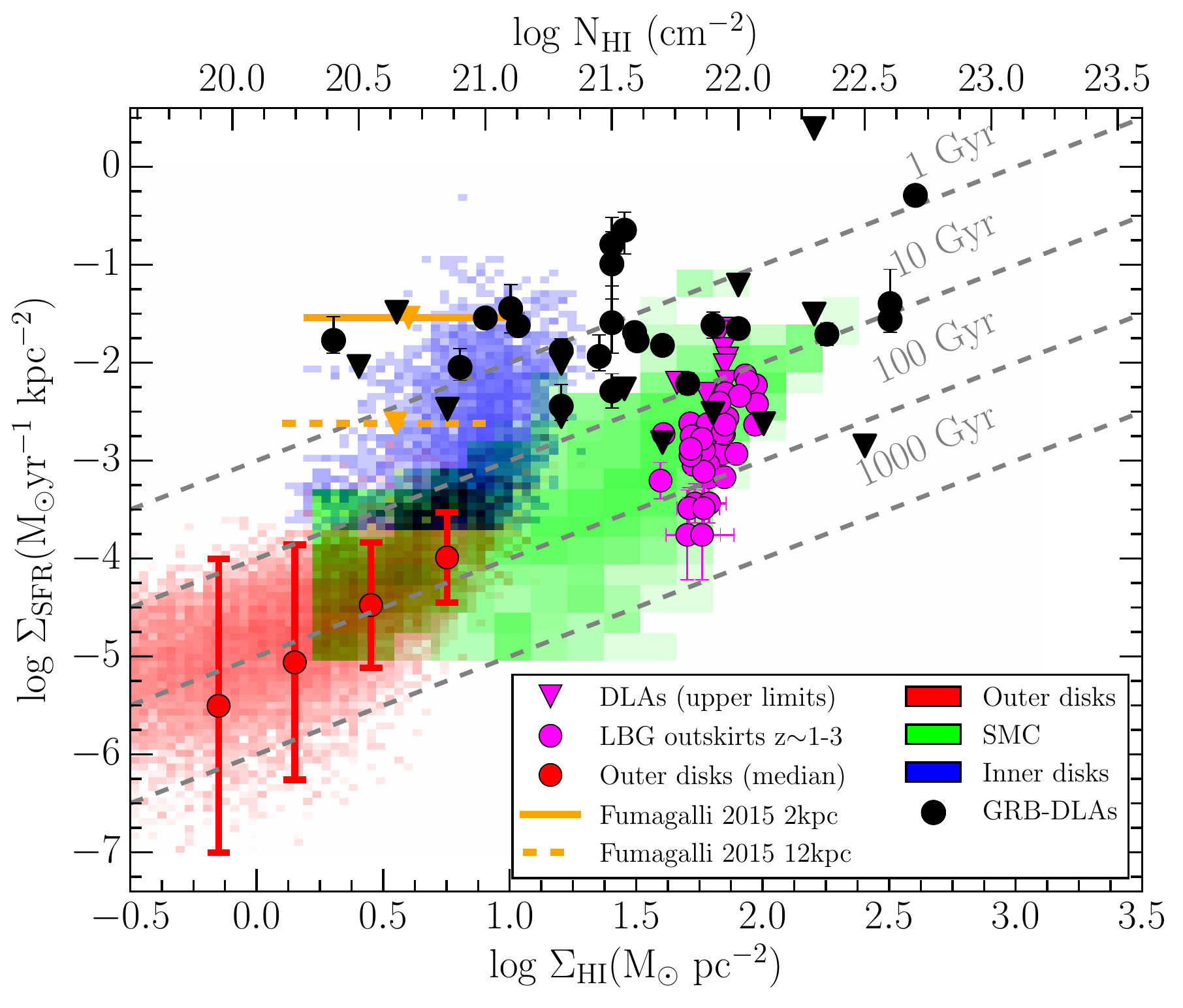}
\caption{Dust-corrected SFR surface densities vs. HI gas surface densities from our GRB-DLAs (black) on top of the compilation of observed SFR surface density vs. gas surface density from \citet{Krumholz:2014}.  The purple points are Lyman break galaxies outskirts at $z \sim 1-3$ (uncorrected for dust) from \citet{Rafelski:2011, Rafelski:2016}, the purple triangles are DLA limits from \citet{Wolfe:2006}, the orange triangles with error bars are composite image limits from dust-uncorrected double-DLAs from \citet{Fumagalli:2015}.  The red pixels are from lines-of-sight through the outer disks of local spiral and dwarf galaxies \citep{Bigiel:2010}.  The red circles are the median and $1-\sigma$ scatter.  The blue pixels are from the inner parts of local galaxies \citep{Bigiel:2008}.  The green pixels are from the SMC \citep{Bolatto:2011}.  Note the SMC, LBG outskirts, and DLA limits are actually plotted for the SFR surface densities vs. the {\it total} neutral hydrogen gas surface densities, not the {\it atomic} hydrogen gas surface density.  We expect that adjusting these measurements to the HI gas surface densities will move the points to the left. The gray dashed lines mark constant depletion times ($t_{\rm dep} = M_{\rm gas}/ \rm SFR$).}
\label{fig:surfacedenscomp}
\end{figure*}


\subsection{Kennicutt-Schmidt relation}
\label{sec:KS}

The Kennicutt-Schmidt relation (KS-relation) connects the available neutral hydrogen gas surface density to form stars ($\Sigma_{\rm HI}$) to the actual measured star formation rate surface density ($\Sigma_{\rm SFR}$).  The KS-relation
has been extensively studied in the local Universe \citep{Bigiel:2008,Bigiel:2010,Bolatto:2011,Elmegreen:2015}. As we mentioned previously we only consider the atomic hydrogen gas content since the molecular hydrogen gas has a negligible contribution at these HI column densities. This scenario may change with redshift, metallicity, or the actual regions in which the SFR is measured - core vs. outskirts of galaxies \citep[e.g.][]{Glover:2012,Krumholz:2012,Krumholz:2013,Rafelski:2016}.


\subsubsection{Surface Density Estimates}
\label{sec:calsurfdens}

In order to measure $\Sigma_{\rm HI}$, which is estimated \emph{along the line-of-sight of the GRB}, we 
assume that the neutral gas is equally distributed across the entire PSF used for determining our SFR density. 
Figure\,\ref{fig:cartoon} shows the idealized case (panel 1a) where the GRB is well localized and the host galaxy 
is resolved.  We include an observed example of this idealized case (panel 1b) for the DLA galaxy identified in 
the \hst\ image of GRB\,050820A (cigar shaped with bright nucleus to the south; see \citealt{Blanchard:2016} for compilation of GRB host galaxy morphologies): the GRB location is identified with sub-arcsec precision due to rapid follow-up of the afterglow with \hst\, (red circle in 1a and 1b panels), and is in the outskirt of the host galaxy (at radius $r_{HI}$).
Moreover, the \nhi\ column is measured through the same environment (which may vary at smaller impact 
parameters).

However, in general, due to the high-redshift nature and the quality of our data, we encounter a less ideal 
scenario, as shown in Figure\,\ref{fig:cartoon} panels 2a and 2b. The uncertainty in the GRB localization, 
despite being often $\lesssim 1\arcsec$ ($1\arcsec$ is $\sim$6-9 kpc for $z =$ 2-6), combined with the unresolved host morphology do not allow us
to accurately measure $\Sigma_{\rm HI}$ and $\Sigma_{\rm SFR}$. In particular, as evident in panel 2a, the 
uncertainty in the GRB localization (red circle) makes it difficult to determine the actual neutral hydrogen line-of-sight 
($r_{1,HI}$ and $r_{2,HI}$ are equally viable, but clearly probe two very different environments). 

In order to be consistent with the local observed KS-relation and the higher-$z$ theoretical models, 
we calculate $\Sigma_{\rm SFR}$ using our dust-corrected SFR calculated in \S\ref{sec:SFR} and the 
area covered by the unresolved ground-based aperture (1.2\arcsec\, radius aperture) around the GRB location, 
which correspond to a circular area of $\sim$17 kpc diameter for $z = 2-6$ (astropy's FlatLambdaCDM; \citealt{Astropy:2013}), for all the objects in our sample.  While this area decreases 
the $\Sigma_{\rm SFR}$ for our resolved \hst\ objects by a factor of $\sim$70, this allows us to be consistent when we compare both our 
resolved and unresolved observations to other samples and models.

Furthermore, in this context we derive the atomic gas (HI) surface density, $\Sigma_{\rm HI}$, directly from 
the DLA line-of-sight neutral hydrogen column density as shown by \citet{Lanzetta:2002aa} and \citet{Hopkins:2005aa} 
even though $\Sigma_{\rm HI}$ and $\Sigma_{\rm SFR}$ are measured over different scales, the KS-relation is, on 
average, still  valid \citep[see also, e.g.,][for the limitations of such approximation]{Zwaan:2006aa,Wolfe:2006aa,Rafelski:2011}.
This is clearly an oversimplification, but it is consistent with the analyses from other SF laws and cosmological simulations. Note 
that we do not include GRB-DLAs that have no $A_V$ measurements as the dust-corrected SFR 
measurements are usually lower limits.


\subsubsection{Comparision with star formation laws \& simulations}

With these caveats in mind, and in order to be consistent with previous works, we overplot the local Kennicutt-Schmidt relation of $\Sigma_{\rm SFR} = K (\frac{\Sigma_{\rm HI}}{\Sigma_{0}})^{\beta}$ with K = $(2.5 
\pm 0.7) \times 10^{-4} \rm M_{\odot} \rm yr^{-1} \rm kpc^{-2}$, $\beta = 1.40 \pm 0.15$, and $\Sigma_0 = 1 \rm 
M_{\odot} \rm pc^{-2}$ \citep{Kennicutt:1998} in Figure \ref{fig:surfacedens} along with a dynamical star 
formation law for spiral and irregular galaxies \citep{Elmegreen:2015} and a star formation law at $z \sim 3$ 
from cosmological simulations \citep{Gnedin:2010}.   Note that the \citet{Gnedin:2010} SFR surface density is 
shown for the total neutral hydrogen gas (dash blue line), only molecular hydrogen gas (dotted-dash blue line), 
and only atomic hydrogen gas (solid blue line). We also, in the two panels, color code our points based on GRB 
afterglow absorption-line metallicity (left) and redshift (right). 
The interpretation of this plot is clearly non-trivial: a large fraction ($\sim$50\%) of our detected DLA counterpart falls in the predicted local K-S relation (shaded area), while some very low metallicity systems are below. Moreover, the presence of our upper limits seem to indicate a very low $\Sigma_{\rm SFR}$ for the amount of
measured $\Sigma_{\rm HI}$. These discrepancies can be due to different factors: GRB afterglow measured metallicities may be lower than the average DLA-host metallicity or the distribution of neutral hydrogen may be 
poorly approximated \citep{Lanzetta:2002aa,Hopkins:2005aa}.
Finally, while we emphasize here that most of our metallicity estimates are lower limits, the $z \sim 3$ theoretical
predictions seem to better predict some of the low metallicity and high metallicity systems.

In Figure \ref{fig:surfacedenscomp} we overlay our DLA counterparts onto results from \citet{Krumholz:2014} 
showing the star formation efficiencies in LBG outskirts \citep{Rafelski:2011, Rafelski:2016}, previous DLA upper 
limits \citep{Wolfe:2006}, double-DLA composite image limits \citep{Fumagalli:2015}, the outer disks of local spiral and dwarf galaxies using 21cm emission to measure HI \citep{Bigiel:2010}, the inner disks of the local using 21cm emission to measure HI \citep{Bigiel:2008}, and the 
Small Magellanic Cloud (SMC; \citealt{Bolatto:2011}).  \citet{Rafelski:2011, Rafelski:2016} report dust-uncorrected $\Sigma_{\rm SFR}$ and use a different SFR conversion (see \S \ref{sec:singleSFR}) which may partially explain our higher $\Sigma_{\rm SFR}$ for GRB-DLA hosts (although
some discrepancies may still remain). We note that \citet{Krumholz:2014} originally plotted the 
SFR surface density against {\it total} neutral hydrogen gas surface density not the {\it atomic} hydrogen gas 
surface density. We expect that adjusting these measurements to HI gas surface densities will shift the magenta points to the left in the plot. 

We also overplot lines of constant depletion times, $t_{\rm dep} = M_{\rm gas}/ \rm SFR$.  Depletion time 
represents how long it would take to completely use up the neutral gas (in this case, HI) with a constant SFR. Our sample covers a large range of depletion times, some of which are longer than the age of the Universe as seen by the galaxy at the DLA redshift.  
This indicates that some of these systems have not reached equilibrium yet and that we are measuring a phase of lower star-formation than in 
earlier times. 

Our GRB-DLA counterparts seem to show no overlap with local outer disk galaxies and seem to have similar 
depletion times as inner galaxy disks, the SMC, and LBG outskirts (in a few cases).  
This is consistent with the observational evidence that GRB hosts are compact, SMC type, star-forming  
galaxies (see also \citealt{Noterdaeme:2012a}). However, we would caution the reader that GRB-DLAs may 
sample higher SFRs than QSO-DLAs because GRBs are associated with massive stars that are typically in  
galaxies with higher specific SFRs. From the DLA counterpart perspective this shows that our sample traces DLAs with 
shorter depletion times than other DLAs or LBG outskirts (magenta dots; note that these points are dust-uncorrected SFRs), and that its higher metallicity, typically 
1\%-20\% the solar value,  can be the cause of this offset \citep[see][]{Krumholz:2014}. For the same reason, 
most of the magenta points in Figure \ref{fig:surfacedenscomp} have much longer depletion times at fixed gas 
surface density than most local spirals.

\begin{figure*}[ht]
\centering
\includegraphics[width=4in]{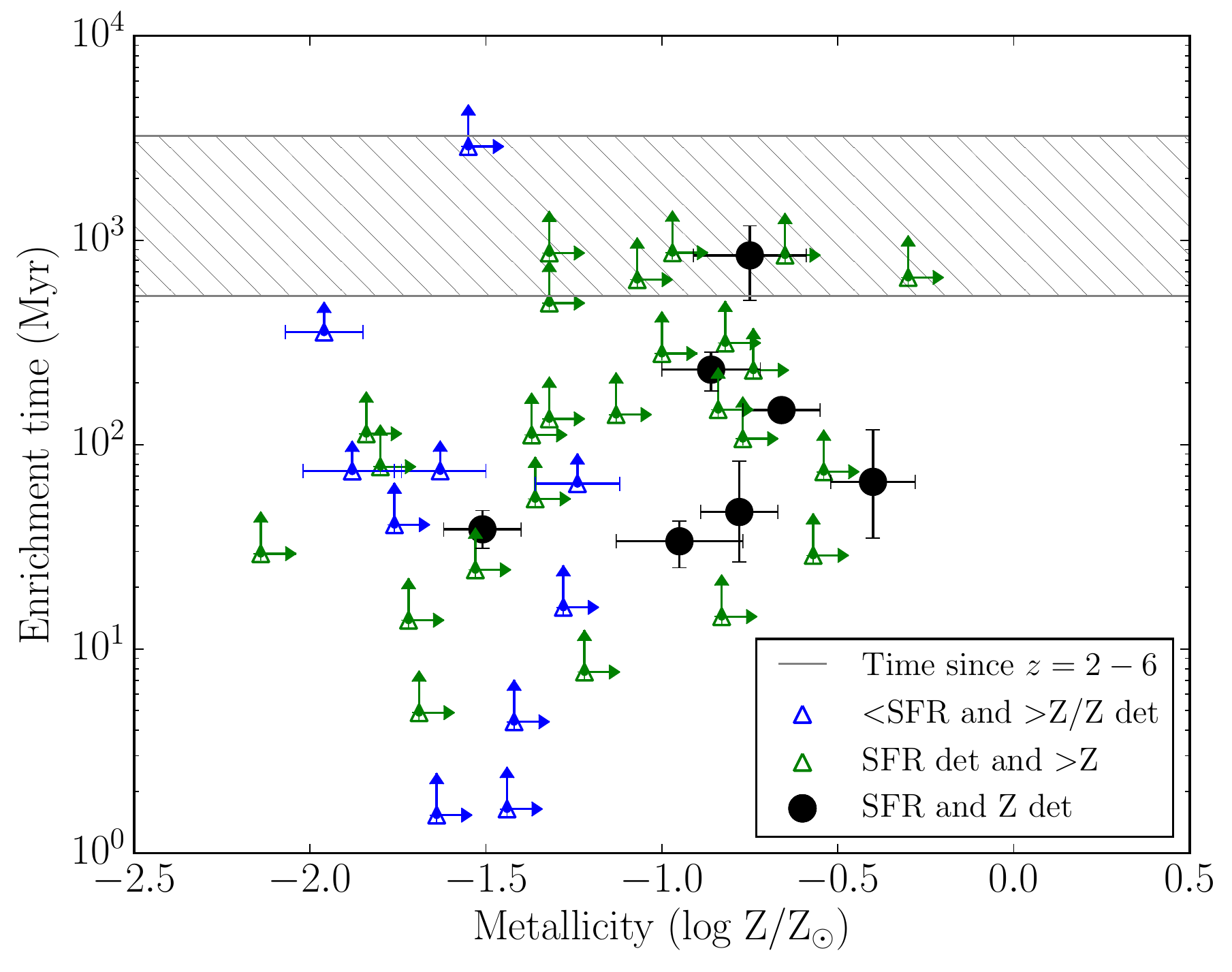}
\caption{ Enrichment time assuming that the galaxy has maintained a constant SFR and that the observed absorption-line metallicity is the same as the galaxy-wide metallicity which is purely determined by internal star formation activity.
We only have lower limits or detections for metallicity and upper limits and detections for SFRs.  Limits are plotted with triangles and the colors represent detected SFRs (green) or upper limit SFRs (blue). Black circles have measured metallicity and measured SFR.  The hatched gray area is the time from z=10 to our DLA redshifts of 2-6.}
\label{fig:enrich}
\end{figure*}


\section{Enrichment time}

Star formation is only process responsible for metal production. Supernova feedback and stellar winds, on 
the other hand, contribute to the dispersion of metals towards the outer regions or even outside the galaxy's 
potential well. The enrichment time is used to determine if the current star formation rate can solely account for 
the current measured metallicity and the metal build up of these systems. We assume a very simple scenario 
where the star formation rate is constant and the metal mass is calculated from the absorption-line metallicity 
measured from GRB afterglow spectra (``closed box'' model).

We calculate the mass in metals:

\begin{equation}
\label{eq:massmet}
M_{\mathrm{z, obs}} = 10^{[X/H]} \rm Z_{\odot} m_p N_{HI} \pi r^2 
\end{equation}
where [X/H] is the metallicity measured from absorption listed in Table \ref{tab:DLAmaster}, assuming Z$_{\odot} = 0.0181$ \citep{Asplund:2009} and $r$ is the radius that we take to be 1.2" across all redshifts.  We then assume that the observed mass in metals is solely due to star formation and we can calculate the enrichment time, $\Delta t_z$, from 

\begin{equation}
\label{eq:enrich}
M_{\mathrm{z, SFR}} = y_z \dot{\psi} \Delta t_z
\end{equation}
where we assume a metal yield of $y_z$ = 1/42 \citep{Madau:1996} and use SFRs ($\dot{\psi}$) from Table \ref{tab:DLAmaster}.  Note that Eq. \ref{eq:massmet} may overestimate the mass of metals particularly because the metals could be not fully mixed and absorption features typically arise in highly enriched gas.  This may lead to inflated enrichment times.

We plot enrichment time against metallicity (Figure \ref{fig:enrich}) and overplot the time since $z = 10$ to 
$z=2$ and $z=6$ (where most of our DLAs are found).  Some of our DLA counterparts have enrichment times shorter than the age of their host galaxy.  This indicates that these galaxies have an underabundance of metals if the metals were formed from a constant 
SFR.  Therefore, it suggests that these systems could have gone through episodic star formation or that 
feedback expelled metals from the galaxy (stellar or supernova feedback;  \citealt{Dave:2007,Rahmati:2016}). On the other hand, other DLA hosts have enrichment times longer than the age of the galaxy.  This means there is an overabundance of metals if the metals were 
formed from a constant SFR.  This may be evidence of either episodic or exponentially declining star formation, poor mixing between the metals within the DLA and the rest of the host galaxy, or another source of metal enrichment such as an influx of metal-enriched gas 
from galaxy mergers. The former have been also invoked by \citet{Hunt:2014}, which has shown that a significant amount of the total 
stellar mass ($\geq 10\%$) of some GRB host galaxies can be created in very short ($\sim 50$Myr) star formation episodes.


\section{Summary and Conclusions}
\label{sec:summary}

We present a sample of 45 DLA galaxy counterparts from photometric follow-up of the GRB host locations.  We 
use a sample of spectroscopically confirmed GRB-DLAs identified in \citet{Cucchiara:2015} and collect all the publically 
available GRB host galaxy photometry.  We supplement these observations with DCT-LMI photometric follow-up.  We present 33 DLA galaxy counterpart detections (though 5 only have $A_V$ upper limits) and 12 upper limits.  This quadruples the number of 
detected DLA counterparts known to date (previously 13, all of which are QSO-DLAs).  These GRB-DLAs have a wider 
range of HI column densities than QSO-DLAs because they are likely located at much smaller impact 
parameters than QSO-DLA host galaxies.  

Our rest-frame UV SFRs are usually
higher than QSO-DLA \emph{in situ} identified using the double-DLA technique \citep{Fumagalli:2015} and, 
while long GRBs come from high SFR areas within their galaxies, we still have upper limits that 
are consistent with the double-DLA sample as well as other DLA surveys \citep[see Table 2 in][]{Fumagalli:2015}.  
From our sample, the SFR does not seem to be correlated with 
either redshift or column density, and we cannot determine if SFR correlates with DLA metallicity due to the 
effect of line saturation and blending in GRB afterglow spectra.

We investigate how our sample relates to the Kennicutt-Schmidt relation by looking at the relationship between 
star formation surface density and HI column density.  Our GRB-DLA galaxy counterpart sample spans both 
high and lower efficiency of star formation compared to a variety of star formation laws (local Kennicutt-
Schmidt relation;  \citealt{Schmidt:1959, Kennicutt:1998}, \citealt{Elmegreen:2015} SF laws, and 
\citealt{Gnedin:2010} simulations at $z\sim3$).  We also compare our sample to objects in the local Universe 
and find that our sample is not consistent with the star formation efficiencies of local spiral and dwarf 
galaxies.  Instead, we find similar efficiencies to local Universe inner disks, SMC, and LBG outskirts, 
complementing what has been currently observed from QSO-DLA counterparts.  We caution the reader that 
our SFRs represent a measurement performed over the integrated host galaxies light while the HI column densities 
are measured locally along the line-of-sight of the GRB afterglows and may be subject to observational biases (metal rich, star-forming
environments) compared to the average HI column density of the DLAs.

We also examine the depletion times of our systems.  Depletion time is a measure of how long it would take to 
completely deplete the DLA gas, HI gas in our case, assuming that the current SFR remains constant.  Our 
sample spans a large range of depletion times (1-100 Gyr). Some of the our sample's depletion times are longer than the current age of the Universe as seen by the galaxy which indicates that these systems have not reached equilibrium 
yet.  

Finally, we investigate the enrichment time of our DLA host counterparts.  Enrichment time is the measure of 
how long it would take to form all the current metals assuming they were solely formed from star formation at 
the current constant SFR.  Some DLA counterparts have enrichment times that are much shorter than the age 
of the galaxy which indicates that the galaxy underwent episodic star formation.  Some DLA counterparts have 
enrichment times that are longer than the age of the galaxy which indicate an overabundance of metals 
assuming a constant SFR.  This suggests that these galaxies may have had episodic star formation histories, 
there may be other sources of metal enrichment such as galaxy mergers, or that there is poor metal mixing 
between the metals in the DLA and the rest of the host galaxy.

The higher detection rate of GRB-DLA host galaxies and their properties (e.g. SFR, metallicity) may 
indicate that QSO-DLAs are an entirely different population than GRB-DLAs.  While investigation of this issue is beyond the scope of this study, 
we note that such a difference may be due to an intrinsic bias in the GRB-DLA sample such that they represent actively star-forming 
regions with special conditions correlated with the likelihood of GRB appearance (e.g. trace different physical regions of galaxy).  Additionally, metallicity may affect the GRB environment differently than QSO-DLAs.

GRB-DLAs are unique objects that have good localization and can later be followed up with photometry and 
spectroscopy. These are key advantages with respect to the identification of DLAs along
QSOs. However, it is unclear if these objects are from the same DLA population.  Our sample, complementary to the QSO-DLAs, is the largest collection of DLA galaxy counterparts available to date bringing the 
total number of detected DLA counterparts from 13 to 58.  Future deep, multi-band, follow-up observations of the remaining GRB-DLAs, in 
particular with \hst\ and large aperture telescopes, will increase the sample size for comparisons with cosmological simulations. 
Furthermore, we showed the importance of accurate identification (sub-arcsecond or 
better) of GRB afterglows in precisely pinpointing the DLA location within their host, especially in lieu of 
more powerful, parsec scale, simulations.
Finally, it will be important to investigate the morphology of DLA hosts, in particular using GRB host galaxies, which seem to show signs of 
pair interaction (Cooke in prep, private communication) and may open new insights on the nature of DLAs and the \emph{in situ} star-formation.


\section*{Acknowledgements}

These results made use of Lowell Observatory's Discovery Channel Telescope.
Lowell operates the DCT in partnership with Boston University, Northern Arizona University, the University of 
Maryland, and the University of Toledo.  Partial support of the DCT was provided by Discovery Communications.  
LMI construction was supported by a grant AST-1005313 from the National Science Foundation.

We gratefully acknowledge M. Krumholtz for sharing his data which significantly improved the analysis for this 
paper.

This work was supported by the National Aeronautics and Space
Administration (NASA) Headquarters under the NASA Earth and Space
Science Fellowship Program (Grant NNX12AL70H to VT). VT, JC, and
SV were partially supported by NSF/ATI grant 1207785.
AC is funded by the NASA grant ``Multiband Observations of the Most Relativistic Gamma-Ray Bursts", NNX15AP23G.
MR acknowledges support from the NASA Postdoctoral Program.
MF acknowledges support by the Science and Technology Facilities Council (Grant number ST/L00075X/1).

\bibliographystyle{apj}
\bibliography{References/dla}


\begin{center}
\begin{deluxetable*}{lllrcllrl}
\tabletypesize{\footnotesize}
\tablecolumns{9}
\tablewidth{0pt}
\tablecaption{GRB-DLAs}
\tablehead{
\colhead{GRB-DLA} & \colhead{Redshift$^a$}  & \colhead{log N$_{\mathrm{HI}}$} & \colhead{log Z/Z$_{\odot}$} & \colhead{$A_V$}
& \colhead{SFR$^b$}& \colhead{log M$_{*}$ } & \colhead{log M$_{\mathrm{dust}}$ } & \colhead{Ref.}\\
&	&\colhead{(cm$^{-2}$)}	&  & &  \colhead{(M$_\odot$ yr$^{-1}$)} &  \colhead{(M$_\odot$)} & \colhead{(M$_\odot$)} & 
}
\startdata
000926	&	2.03621	&	$21.30\pm0.25$	&	$>-0.30$	&	$0.038$	&	$3.03^{+0.97}_{-0.36}$	&	$9.90^{+0.16}_{-0.22}$	&	$6.00^{+0.55}_{-0.00}$	&	(1)	\\
011211	&	2.1427	&	$20.40\pm0.20$	&	$>-1.22$	&	$0.138$	&	$3.86^{+2.85}_{-1.00}$	&	$8.94^{+0.18}_{-0.28}$	&	$6.19^{+0.70}_{-0.19}$	&	(2)	\\
020124	&	3.198	&	$21.70\pm0.20$	&	\nodata	&	$0.280$$\pm 0.330$$^c$	&	$<0.35$	&	\nodata	&	\nodata	&	(3),(4)	\\
030226	&	1.98	&	$20.50\pm0.30$	&	$>-1.28$	&	$0.060$$\pm 0.060$$^c$	&	$<2.08$	&	\nodata	&	\nodata	&	(4),(5)	\\
030323	&	3.3714	&	$21.90\pm0.07$	&	$>-1.32$	&	$<$$0.020$$^c$	&	$<0.69$	&	\nodata	&	\nodata	&	(3),(4)	\\
030429	&	2.658	&	$21.60\pm0.20$	&	$>-1.13$	&	$0.400$$\pm 0.100$$^c$	&	$3.82^{+0.79}_{-0.56}$$^d$	&	\nodata	&	\nodata	&	(3),(4)	\\
050319	&	3.24	&	$20.90\pm0.20$	&	$>-0.77$	&	$0.050$$\pm 0.060$$^c$	&	$2.05^{+1.14}_{-0.54}$$^d$	&	\nodata	&	\nodata	&	(6),(7)	\\
050401	&	2.899	&	$22.60\pm0.30$	&	$>-1.07$	&	$0.738$	&	$9.16^{+11.35}_{-4.52}$	&	$9.56^{+0.23}_{-0.21}$	&	$7.00^{+0.64}_{-0.63}$	&	(3),(6),(8),(9)	\\
050730	&	3.96723	&	$22.10\pm0.10$	&	$-1.96 \pm  0.11$	&	$0.120$$\pm 0.020$$^c$	&	$<0.54$	&	\nodata	&	\nodata	&	(9),(10)	\\
050820A	&	2.6145	&	$21.10\pm0.10$	&	$-0.78 \pm  0.11$	&	$0.813$	&	$8.17^{+6.19}_{-3.57}$	&	$9.16^{+0.17}_{-0.17}$	&	$6.94^{+0.60}_{-0.59}$	&	(3),(6),(8),(9)	\\
050904	&	6.26	&	$21.30\pm0.20$	&	$>-1.00$	&	$<$$0.050$$^c$	&	$<0.64$$^d$	&	\nodata	&	\nodata	&	(11),(12)	\\
050922C	&	2.1996	&	$21.55\pm0.10$	&	$-1.88 \pm  0.14$	&	$0.090$$\pm 0.030$$^c$	&	$<1.23$	&	\nodata	&	\nodata	&	(3),(13)	\\
060115	&	3.533	&	$21.50\pm0.10$	&	$>-1.53$	&	$0.763$	&	$5.85^{+8.02}_{-2.98}$	&	$9.33^{+0.20}_{-0.28}$	&	$6.81^{+0.64}_{-0.65}$	&	(6),(8),(9)	\\
060206	&	4.048	&	$20.85\pm0.10$	&	$>-0.74$	&	$<$$0.170$$^c$	&	$<0.77$	&	\nodata	&	\nodata	&	(9),(13)	\\
060210	&	3.913	&	$21.55\pm0.15$	&	$>-0.83$	&	$0.363$	&	$51.52^{+27.36}_{-22.21}$	&	$9.99^{+0.15}_{-0.12}$	&	$7.52^{+0.61}_{-0.65}$	&	(6),(14)	\\
060223A	&	4.41	&	$21.60\pm0.10$	&	$>-1.80$	&	\nodata	&	$1.03^{+0.24}_{-0.16}$$^d$	&	\nodata	&	\nodata	&	(9)	\\
060510B	&	4.94	&	$21.30\pm0.10$	&	$>-0.84$	&	$<$$0.500$$^c$	&	$<2.23$$^d$	&	\nodata	&	\nodata	&	(6),(14)	\\
060522	&	5.11	&	$21.00\pm0.30$	&	\nodata	&	\nodata	&	$<2.96$	&	\nodata	&	\nodata	&	(15)	\\
060707	&	3.425	&	$21.00\pm0.20$	&	$>-1.69$	&	$0.080$$\pm 0.020$$^c$	&	$6.54^{+0.37}_{-0.35}$	&	\nodata	&	\nodata	&	(8),(10)	\\
060714	&	2.711	&	$21.80\pm0.10$	&	$>-0.97$	&	$0.210$$\pm 0.020$$^c$	&	$1.40^{+0.41}_{-0.32}$	&	\nodata	&	\nodata	&	(8),(10)	\\
060926	&	3.206	&	$22.60\pm0.15$	&	$>-1.32$	&	$0.320$$\pm 0.020$$^c$	&	$6.31^{+3.52}_{-1.67}$$^d$	&	\nodata	&	\nodata	&	(10),(16)	\\
060927	&	5.464	&	$22.50\pm0.15$	&	$>-1.55$	&	$<$$0.170$$^c$	&	$<0.32$	&	\nodata	&	\nodata	&	(9),(13)	\\
061110B	&	3.433	&	$22.35\pm0.10$	&	$>-1.84$	&	$0.230$$\pm 0.030$$^c$	&	$4.46^{+1.37}_{-1.05}$	&	\nodata	&	\nodata	&	(8),(10)	\\
070110	&	2.351	&	$21.70\pm0.10$	&	$>-1.32$	&	$0.100$$\pm 0.100$$^c$	&	$3.43^{+0.63}_{-0.46}$$^d$	&	\nodata	&	\nodata	&	(8),(10)	\\
070506	&	2.308	&	$22.00\pm0.30$	&	$>-0.65$	&	$0.440$$\pm 0.050$$^c$	&	$5.09^{+0.98}_{-0.71}$$^d$	&	\nodata	&	\nodata	&	(8),(10)	\\
070721B	&	3.628	&	$21.50\pm0.20$	&	$>-2.14$	&	$0.200$$\pm 0.020$$^c$	&	$1.17^{+0.59}_{-0.39}$	&	\nodata	&	\nodata	&	(8),(10)	\\
070802	&	2.455	&	$21.50\pm0.20$	&	$>-0.54$	&	$0.838$	&	$23.28^{+26.49}_{-13.12}$	&	$9.71^{+0.11}_{-0.11}$	&	$7.33^{+0.65}_{-0.67}$	&	(8),(9),(17),(18)	\\
080210	&	2.641	&	$21.90\pm0.10$	&	$>-1.37$	&	$0.330$$\pm 0.030$$^c$	&	$5.53^{+1.96}_{-1.45}$	&	\nodata	&	\nodata	&	(5),(10)	\\
080607	&	3.037	&	$22.70\pm0.15$	&	$>-1.72$	&	$2.938$	&	$116.68^{+0.00}_{-0.00}$	&	$10.13^{+0.00}_{-0.00}$	&	$8.36^{+0.49}_{-0.52}$	&	(6),(9),(17)	\\
080804	&	2.20542	&	$21.30\pm0.10$	&	$-0.75 \pm  0.16$	&	$0.170$$\pm 0.110$$^c$	&	$0.82^{+0.54}_{-0.23}$$^d$	&	\nodata	&	\nodata	&	(6),(13)	\\
081008	&	1.96	&	$21.59\pm0.10$	&	$-0.86 \pm  0.14$	&	$0.290$$\pm 0.070$$^c$	&	$4.64^{+1.27}_{-0.82}$$^d$	&	\nodata	&	\nodata	&	(7),(9)	\\
090205	&	4.64	&	$20.73\pm0.05$	&	$>-0.57$	&	\nodata	&	$6.14^{+0.59}_{-0.54}$	&	\nodata	&	\nodata	&	(19)	\\
090516	&	4.109	&	$21.73\pm0.10$	&	$>-1.36$	&	\nodata	&	$5.87^{+3.43}_{-2.17}$	&	\nodata	&	\nodata	&	(20)	\\
090812	&	2.425	&	$22.30\pm0.10$	&	$>-1.64$	&	$0.230$$\pm 0.080$$^c$	&	$<561.26$	&	\nodata	&	\nodata	&	(5),(13)	\\
100219A	&	4.667	&	$21.13\pm0.12$	&	$-0.95 \pm  0.18$	&	$0.130$$\pm 0.050$$^c$	&	$5.45^{+1.89}_{-1.12}$$^d$	&	\nodata	&	\nodata	&	(21)	\\
110205A	&	2.214	&	$21.45\pm0.20$	&	$>-0.82$	&	$0.350$$\pm 0.060$$^c$	&	$2.65^{+1.74}_{-0.75}$$^d$	&	\nodata	&	\nodata	&	(6),(13)	\\
111008A	&	4.98968	&	$22.30\pm0.06$	&	$-1.63 \pm  0.13$	&	$0.110$$\pm 0.040$$^c$	&	$<7.16$	&	\nodata	&	\nodata	&	(22)	\\
120327A	&	2.813	&	$22.01\pm0.09$	&	$-1.51 \pm  0.11$	&	$<$$0.030$$^c$	&	$<14.16$	&	\nodata	&	\nodata	&	(23)	\\
120716A	&	2.487	&	$21.55\pm0.15$	&	$>-1.76$	&	\nodata	&	$<2.84$	&	\nodata	&	\nodata	&	(5)	\\
120909A	&	3.9293	&	$21.20\pm0.10$	&	$-0.66 \pm  0.11$	&	\nodata	&	$3.31^{+0.39}_{-0.35}$	&	\nodata	&	\nodata	&	(20)	\\
121024A	&	2.2977	&	$21.50\pm0.10$	&	$-0.40 \pm  0.12$	&	$0.563$	&	$36.90^{+32.60}_{-16.39}$	&	$10.15^{+0.16}_{-0.17}$	&	$7.54^{+0.60}_{-0.60}$	&	(24)	\\
121201A	&	3.385	&	$21.70\pm0.20$	&	\nodata	&	\nodata	&	$6.45^{+1.38}_{-1.13}$	&	\nodata	&	\nodata	&	(20)	\\
130408A	&	3.757	&	$21.70\pm0.10$	&	$-1.24 \pm  0.12$	&	\nodata	&	$<6.54$	&	\nodata	&	\nodata	&	(20)	\\
130505A	&	2.2687	&	$20.65\pm0.10$	&	$>-1.42$	&	$<$$0.128$$^c$	&	$<7.47$	&	\nodata	&	\nodata	&	(5),(25)	\\
140423A	&	3.258	&	$20.45\pm0.20$	&	$>-1.44$	&	\nodata	&	$<8.95$	&	\nodata	&	\nodata	&	(5)

\enddata
\tablecomments{$^a$Significant digits of redshift reflect accuracy of measurement. $^b$Dust-corrected (except those without $A_V$ measurements). $^c$Host extinction from GRB afterglow measurements. $^d$Calculates SFR from MAGPHYS SED scaling of photometric detection.  
(1) \citealt{Castro:2003}, (2) \citealt{Fynbo:2003}, (3) \citealt{Chen:2009}, (4) \citealt{Kann:2006}, (5) This work, (6) \citealt{Perley:2015a}, (7) \citealt{Schady:2012}, (8) \citealt{Hjorth:2012}, (9) \citealt{Blanchard:2016}, (10) \citealt{Zafar:2011}, (11) \citealt{McGuire:2015}, (12) \citealt{Zafar:2010}, (13) \citealt{Covino:2013}, (14) \citealt{Perley:2009}, (15) \citealt{Basa:2012}, (16) \citealt{Laskar:2011}, (17) \citealt{Perley:2013}, (18) \citealt{Kruhler:2011}, (19) \citealt{Davanzo:2010}, (20) \citealt{Greiner:2015}, (21) \citealt{Thone:2013}, (22) \citealt{Sparre:2014}, (23) \citealt{DElia:2014}, (24) \citealt{Friis:2015}, (25) \citealt{Cannizzo:2013}}
\label{tab:DLAmaster}
\end{deluxetable*}
\end{center}

\clearpage

\begin{center}
\begin{deluxetable*}{lllrcllrl}
\tabletypesize{\footnotesize}
\tablecolumns{9}
\tablewidth{0pt}
\tablecaption{GRB-sub-DLAs}
\tablehead{
\colhead{GRB sub-DLA} & \colhead{Redshift$^a$}  & \colhead{log N$_{\mathrm{HI}}$} & \colhead{log Z/Z$_{\odot}$} & \colhead{$A_V$}
& \colhead{SFR$^b$} & \colhead{log M$_{*}$ } & \colhead{log M$_{\mathrm{dust}}$ } & \colhead{Ref.}\\
&	&\colhead{(cm$^{-2}$)}	&  & &  \colhead{(M$_\odot$ yr$^{-1}$)} &  \colhead{(M$_\odot$)} & \colhead{(M$_\odot$)} & 
}
\startdata
021004	&	2.3289	&	$19.00\pm0.20$	&	\nodata	&	$0.038$	&	$7.19^{+0.17}_{-1.21}$	&	$9.29^{+0.06}_{-0.32}$	&	$9.29^{+0.06}_{-0.32}$	&	(1),(2)	\\
050908	&	3.344	&	$19.40\pm0.20$	&	\nodata	&	$<$$0.550$$^c$	&	$3.21^{+1.65}_{-1.09}$	&	\nodata	&	\nodata	&	(3),(4)	\\
060124	&	2.3	&	$18.50\pm0.50$	&	\nodata	&	$0.170$$\pm 0.030$$^c$	&	$0.46^{+0.10}_{-0.07}$$^d$	&	\nodata	&	\nodata	&	(5),(6)	\\
060526	&	3.221	&	$19.90\pm0.15$	&	\nodata	&	$0.700$$\pm 0.180$$^c$	&	$1.63^{+0.91}_{-0.43}$$^d$	&	\nodata	&	\nodata	&	(6),(7)	\\
060605	&	3.773	&	$18.90\pm0.40$	&	\nodata	&	\nodata	&	$0.40^{+0.06}_{-0.05}$	&	\nodata	&	\nodata	&	(5)	\\
060607A	&	3.075	&	$16.95\pm0.03$	&	\nodata	&	$0.080$$\pm 0.040$$^c$	&	$<0.29$	&	\nodata	&	\nodata	&	(3),(8)	\\
080310	&	2.427	&	$18.70\pm0.10$	&	\nodata	&	$0.100$$\pm 0.020$$^c$	&	$1.82^{+1.14}_{-0.51}$$^d$	&	\nodata	&	\nodata	&	(4),(7)	\\
080810	&	3.35	&	$17.50\pm0.15$	&	\nodata	&	\nodata	&	$27.33^{+15.99}_{-10.09}$	&	\nodata	&	\nodata	&	(9)	\\
080913	&	6.69	&	$<19.84$	&	\nodata	&	$0.120$$\pm 0.030$$^c$	&	$<1.51$	&	\nodata	&	\nodata	&	(10),(11)	\\
090323	&	3.5778	&	$>19.90$	&	\nodata	&	\nodata	&	$9.72^{+1.75}_{-1.49}$	&	\nodata	&	\nodata	&	(12)	\\
090426	&	2.609	&	$19.10\pm0.15$	&	\nodata	&	$0.088$	&	$3.03^{+0.00}_{-0.00}$	&	$8.48^{+0.00}_{-0.00}$	&	$8.48^{+0.00}_{-0.00}$	&	(13)	\\
130606A	&	5.9134	&	$19.93\pm0.20$	&	\nodata	&	\nodata	&	$1.63^{+0.37}_{-0.25}$$^d$	&	\nodata	&	\nodata	&	(14)	

\enddata
\tablecomments{$^a$Significant digits of redshift reflect accuracy of measurement. $^b$Dust-corrected (except those without $A_V$ measurements). $^c$Host extinction from GRB afterglow measurements. $^d$Calculates SFR from MAGPHYS SED scaling of photometric detection.  
(1) \citealt{Fynbo:2005}, (2) \citealt{deUgartePostigo:2005}, (3) \citealt{Hjorth:2012}, (4) \citealt{Perley:2009}, (5) \citealt{Blanchard:2016}, (6) \citealt{Kann:2010}, (7) \citealt{Perley:2015a}, (8) \citealt{Schady:2012}, (9) \citealt{Greiner:2015}, (10) \citealt{Basa:2012}, (11) \citealt{Zafar:2011}, (12) \citealt{McBreen:2010}, (13) \citealt{Thone:2011}, (14) \citealt{McGuire:2015}}
\label{tab:subDLAmaster}
\end{deluxetable*}
\end{center}

\end{document}